\documentclass{elsartm}

\usepackage[dvips]{graphicx}

\usepackage{amssymb}

\def\degree{\kern-.2em\r{}\kern-.1em}

\begin{document}


\begin{frontmatter}

\title{Design of a Nanometer Beam Size Monitor for ATF2}
\author[uticepp]{Taikan Suehara},
\ead{suehara@icepp.s.u-tokyo.ac.jp}
\author[utphys]{Masahiro Oroku},
\author[utphys]{Takashi Yamanaka},
\author[utphys]{Hakutaro Yoda},
\author[utphys]{Tomoya Nakamura},
\author[uticepp]{Yoshio Kamiya},
\author[kekacc]{Yosuke Honda},
\author[kekapp]{Tatsuya Kume},
\author[kekpar]{Toshiaki Tauchi},
\author[tohoku]{Tomoyuki Sanuki}, and
\author[utphys]{Sachio Komamiya}
\address[uticepp]{ICEPP, The Univ. of Tokyo, 7-3-1 Hongo, Bunkyo, Tokyo, 113-0033, Japan}
\address[utphys]{Dept. of Physics, The Univ. of Tokyo, 7-3-1 Hongo, Bunkyo, Tokyo, 113-0033, Japan}
\address[kekacc]{Accelerator Laboratory, KEK, 1-1 Oho, Tsukuba, Ibaraki, 305-0801, Japan}
\address[kekapp]{Applied Research Laboratory, KEK, 1-1 Oho, Tsukuba, Ibaraki, 305-0801, Japan}
\address[kekpar]{Institute of Particle and Nuclear Studies, KEK, 1-1 Oho, Tsukuba, Ibaraki, 305-0801, Japan}
\address[tohoku]{Dept. of Physics, Tohoku Univ., 6-3 Aoba, Aramaki, Aoba, Sendai, Miyagi, 980-8578, Japan}

\title{}

\author{}

\address{}

\begin{abstract}
We developed an electron beam size monitor for extremely small beam sizes.
It uses a laser interference fringe for a scattering target with the electron beam.
Our target performance is $<$ 2 nm systematic error for 37 nm beam size
and $<$ 10\% statistical error in a measurement using 90 electron bunches for 25 - 6000 nm beam size.
A precise laser interference fringe control system using an active feedback function
is incorporated to the monitor to achieve the target performance.
We describe an overall design, implementations, and performance estimations of the monitor.
\end{abstract}

\end{frontmatter}


\section{Introduction}
\label{sec:intro}

\subsection{Principles}

Nanometer focusing of the electron and positron beams is one of the key technologies
to realize the coming International Linear Collider (ILC)\cite{ilcrdr}.
The beam size at the ILC interaction point (IP) is designed to be 640 nm (horizontal) by
5.7 nm (vertical) to achieve the required integrated luminosity of 500 fb$^{-1}$ within
first four years of operation.
To achieve these beam sizes, especially 5.7 nm vertical beam size,
precise tuning of position and field strength for magnets in the final focus line is required.
An IP beam size monitor is necessary for the tuning and for demonstrations of the nanometer focusing.

For intense electron\footnote{Positrons can be treated as similar.}
 beams, movable fine wire targets are widely used to acquire the beam size.
Electron beams scatter with the wire targets, emitting photons which can be captured by gamma detectors.
Metal and carbon wires down to several $\mu$m thickness have been used for the targets.
However, for sub-$\mu$m beam size, electron beams are so intense that they break the wires\cite{fftbwire},
thus material wires cannot be utilized.

To avoid the heat destruction,
laser beams can be alternatives of the material wires for intense electron beams.
Since laser beams scatter with electron beams emitting inverse-Compton scattered photons,
they can be used as similar to material wires (laser-wire).
Minimum observable electron beam size by a laser-wire is determined by the laser spot size,
which is limited to around its wavelength by diffraction limit.
Therefore, sub-100 nm beam size measurement needs deep-UV lasers, which is not available now.

\begin{figure}
\begin{center}
\includegraphics[width=35em]{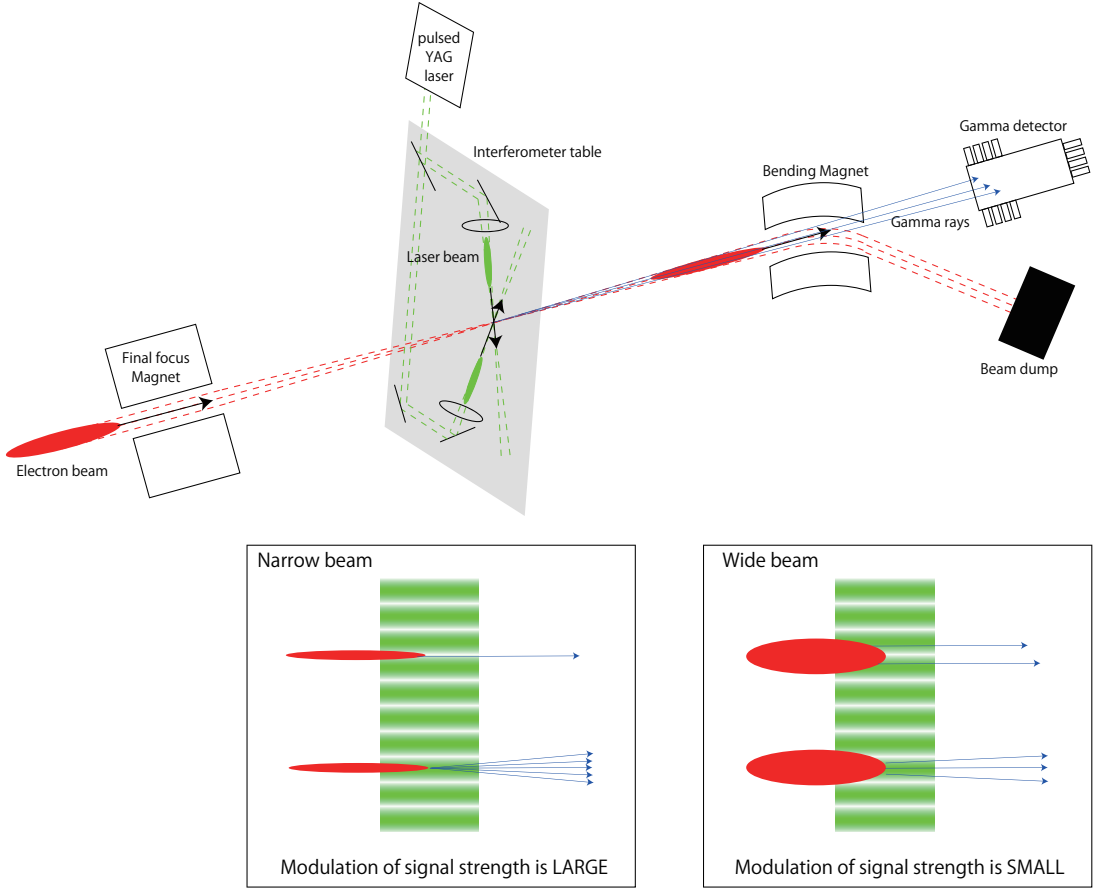}
\caption{A schematic of the laser interferometer (Shintake monitor).}
\label{fig:schematic}
\end{center}
\end{figure}

For sub-100 nm electron beam size,
a laser interferometer technology, called Shintake monitor\cite{shintake1} can be utilized.
Figure \ref{fig:schematic} shows a schematic of the monitor.
A laser beam is split in two and the split beams cross at the focal point
of the electron beam.
In the intersected area of the two laser beams,
the electromagnetic fields of the two laser beams form a standing wave
(interference fringe).
Probability of the Compton scattering varies
by the phase of the standing wave where the electrons pass through,
that is, the scattering probability is high for electrons
passing through the top of the fringe,
and is low for those passing through the bottom of the fringe.
Note that the standing wave of the magnetic and the electric field 
has opposite phase, but for the high energy electron beam,
effect of the electric field is strongly suppressed if we choose
the magnetic field direction
perpendicular to the electron beam direction.
The detailed discussion is appeared in \cite{dron}.

\begin{figure}
\begin{center}
\includegraphics[width=30em]{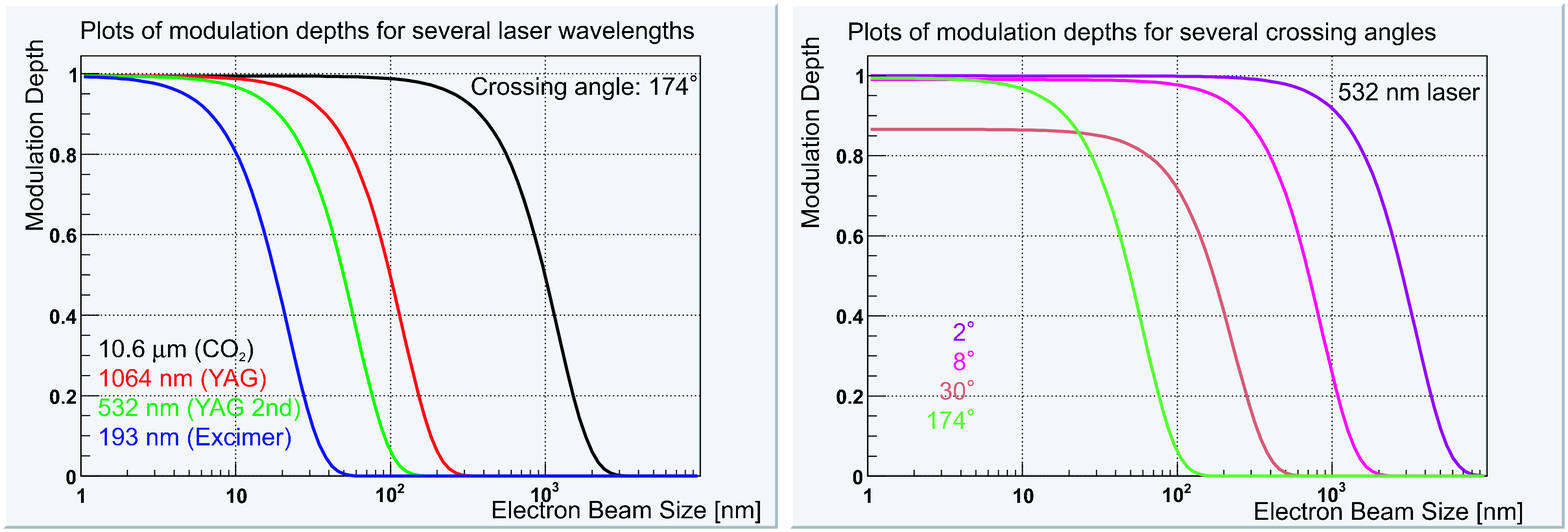}
\caption{Relation between electron beam sizes and modulation depths. Left: comparison between
several laser wavelengths. Right: comparison between several crossing angles.}
\label{fig:moddepth}
\end{center}
\end{figure}

For electron beams well focused compared to the fringe spacing,
all electrons in the beam pass through almost the same phase of the fringe.
This results in a large modulation of the Compton scattering signals
monitored at a gamma detector downstream of the electron beam line.
On the contrary, for dispersed electron beams,
electrons pass through wide variety of phases of the fringe,
thus the modulation of the signals is low.
By calculation of the magnetic field described in \cite{shintake1-2},
electron beam size $\sigma$ is related to the modulation of the monitored Compton signal
$M = |(N_+ - N_-) / (N_+ + N_-)|$ (where $N_+$ is the maximum signal intensity of the modulation
and $N_-$ is the minimum signal intensity of the modulation) as,
\begin{equation}
	M = | \cos2\phi | \exp\left[-2(k_\perp\sigma)^2\right]
	\label{eqn:moddepth}
\end{equation}
where $\phi$ is the crossing angle (half angle) of the two laser beams and
$k_\perp = k\sin\phi$ is the wavenumber along the direction perpendicular to the fringe.
$M = 5-90\%$ can be used for the beam size measurement, thinking of various measurement fluctuation.
Observable beam size range of the monitor is varied by $k_\perp$, which is determined by
laser wavelength and crossing angle, as shown in Fig.~\ref{fig:moddepth}.
Beam sizes down to $<$ 10 nm is observable by this method if a UV laser with a large crossing angle
is selected.

\subsection{Modulation Measurement}

Practically, accelerated electron beam comes in bunches, and we introduce laser pulses to the IP
to interact with the electron bunches.
We obtain Compton signal strength from a bunch (or several bunches) at certain phase of the laser
interference fringe, and obtain signal strength from another bunch (or other bunches) at another phase.
Repeating that results in a modulation spectrum, which is a plot of Compton signal strengths for
fringe phases.

\begin{figure}
\begin{center}
\includegraphics[width=20em]{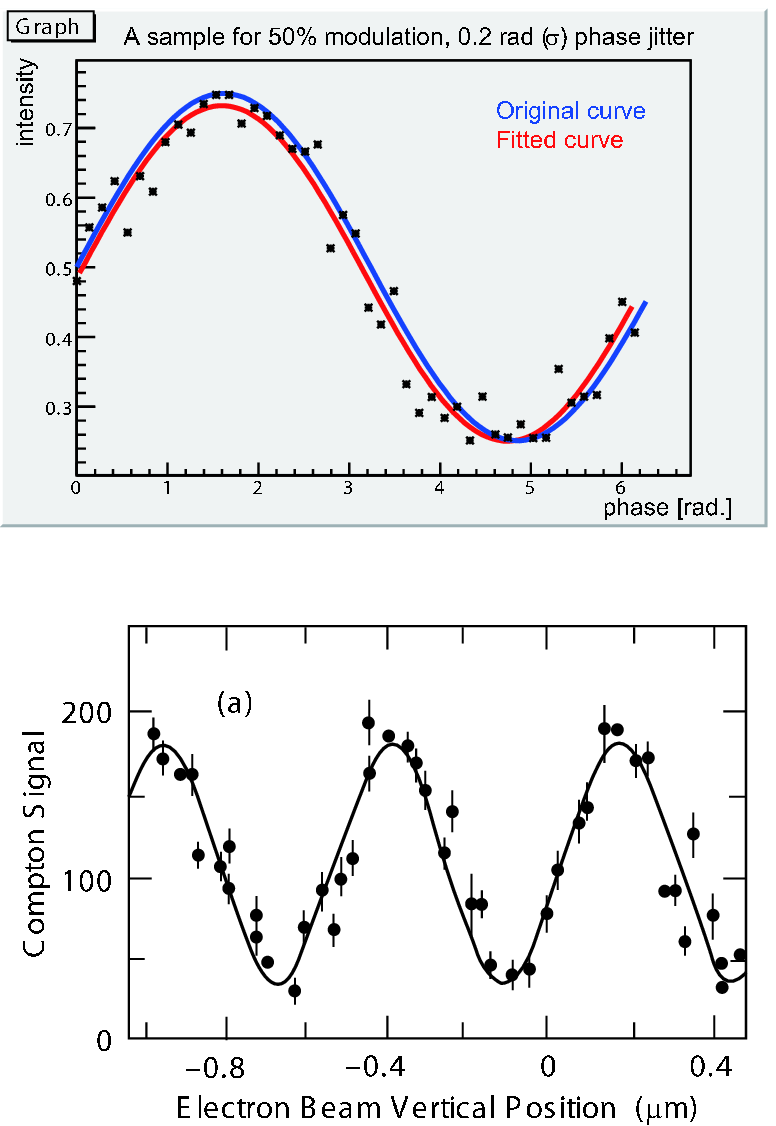}
\caption{Upper: a sample modulation spectrum produced by a toy Monte Carlo simulation.
0.2 radian RMS phase fluctuation is introduced in each point, and the fluctuation causes
deviation of the sine fitting (red line) from the original curve (blue line).
Lower: a modulation spectrum measured by the FFTB Shintake monitor.}
\label{fig:modulation-spec}
\end{center}
\end{figure}

Figure \ref{fig:modulation-spec} (upper) shows a sample modulation spectrum.
Each point stands for a signal intensity in a bunch, including measurement errors
such as power fluctuation, phase fluctuation and background fluctuation.
To obtain good modulation spectrum and thus good resolution of beam size measurements,
we need to suppress these error factors.

Shintake monitor is firstly realized in Final Focus Test Beam experiment\cite{fftb},
constructed in Standford Linear Accelerator Center.
Figure \ref{fig:modulation-spec} (lower) shows an example of the modulation spectrum
obtained in the FFTB Shintake monitor\cite{shintake2}.
In the FFTB Shintake monitor, fringe phase at the electron beam was controlled by shifting
the path of the electron beam by a steering magnet because they had no phase control
feature in their laser optical system.
In our monitor, the laser fringe phase is monitored and controlled instead of shifting electron
beam path, as described in Section \ref{sec:fringe}.


\section{Layout and Structure}

\subsection{Overall Structure}

Our monitor is designed as the IP\footnote{Although ATF2 causes no interaction at the focal point,
we call it `IP' to clarify a relation to the ILC.}
beam size monitor of Accelerator Test Facility 2 (ATF2)\cite{atf2},
a final focus test facility for the ILC.
ATF2 is being constructed downstream an ILC dumping ring test facility, Accelerator Test Facility (ATF)
at High Energy Accelerator Research Organization (KEK).
Major specifications of ATF and ATF2 are shown in Table \ref{tbl:atf2}.
Design beam size at the ATF2 IP is $\sigma_y^\ast = 37$ nm, $\sigma_x^\ast = 2.8$ $\mu$m.
Since the monitor is also used as a beam tuning tool to obtain small beam size,
target beam size of the monitor is 25 nm to 6 $\mu$m for $\sigma_y^\ast$, and 2.8 to 100 $\mu$m for $\sigma_x^\ast$.
We implement a Shintake monitor with several crossing angles for $\sigma_y^\ast$ measurements,
and a laser-wire for $\sigma_x^\ast$ measurements, incorporated in a single IP beam size monitor system.

\begin{table}
	\begin{center}
		\begin{tabular}{|c|c|}\hline
			Beam Energy & 1.3 GeV \\ \hline
			Normalized Emittance $\gamma\epsilon_x$ & $3 \times 10^{-6}$ m$\cdot$rad \\ \hline
			Normalized Emittance $\gamma\epsilon_y$ & $3 \times 10^{-8}$ m$\cdot$rad \\ \hline
			Bunch Population & $0.5 \times 10^{10}$ \\ \hline
			Bunch Length & 5 mm (17 psec) \\ \hline
			Repetition Rate & 1.56 Hz \\ \hline
			Focal Length $L^\ast$ & 1.0 m \\ \hline
			IP Beta Function $\beta_x^\ast$ & 4.0 mm \\ \hline
			IP Beta Function $\beta_y^\ast$ & 0.1 mm \\ \hline
			IP Beam Size $\sigma_x^\ast$ & 2.8 $\mu$m \\ \hline
			IP Beam Size $\sigma_y^\ast$ & 37 nm \\ \hline
		\end{tabular}
	\caption{Major specifications of ATF \& ATF2.}
	\label{tbl:atf2}
	\end{center}
\end{table}

\begin{figure}
\begin{center}
\includegraphics[width=20em]{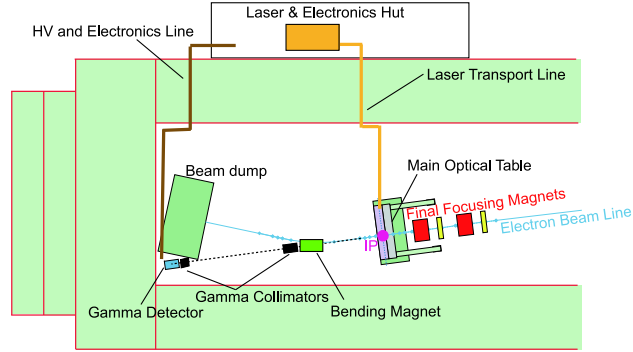}
\caption{Schematic layout of the monitor and around the ATF2 IP.}
\label{fig:layout}
\end{center}
\end{figure}

Figure \ref{fig:layout} shows a schematic layout of the monitor.
It consists of a main optical table, a gamma detector with collimators,
a laser source with a transport line, and control electronics.
The main optical table is installed just on the IP to form laser interference fringes.
Laser photons are supplied from a laser located in the laser \& electronics hut,
transported via the laser transport line.
The electron beam is focused at the IP by the final focusing magnets.
After the IP, the electron beam is bended at the bending magnet and sent to the beam dump.
Compton photons are not bended by the bending magnet, and go straight to the gamma detector
through apertures of the gamma collimators.

\subsection{Laser and Main Optical Table}

To obtain enough number of photons, we need high density of laser photons at the IP.
We use a high power Q-switched pulsed laser whose peak power is up to 40 MW.
Pulse length of the laser is about 8 nsec (FWHM), which is much longer than
the electron bunch length 17 psec.
The laser output is triggered by a electron beam signal observed by a beam position monitor
in the ATF dumping ring.
Q-switch timing jitter of the laser is measured to be about 300 psec,
which causes no significant density flucutation of photons interacted with an electron beam.
Wavelength of the laser is 532 nm (YAG 2nd harmonics) to obtain good sensitivity
for the monitor at 37 nm electron beam size (see Fig.~\ref{fig:moddepth}).
Spectral width of the laser is $< 90$ MHz, narrow enough to obtain good fringe contrast at the IP.

\begin{figure}
\begin{center}
\includegraphics[width=20em]{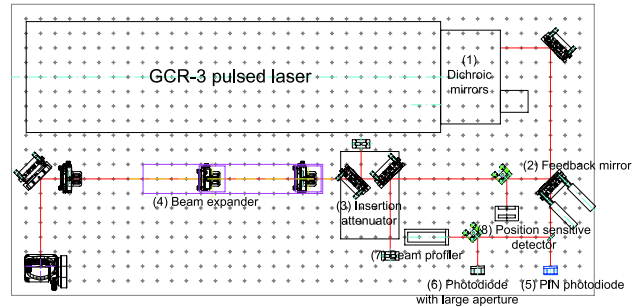}
\caption{Layout of the laser table. Red lines show laser optical paths.}
\label{fig:laser-table}
\end{center}
\end{figure}

Figure \ref{fig:laser-table} shows a layout of the laser table.
The purpose of the table is to adjust and monitor characteristics of the laser beam
to be sent to the main optical table.
A 532 nm monochromatic laser beam is created by the laser and dichroic mirrors(1).
A feedback mirror(2) is a partial($\sim$95\%) reflecting mirror with actuators used for
laser beam position stabilization. An insertion attenuator(3)
can provide low power laser paths for an alignment.
A beam expander(4), which consists of three lenses, magnifies a laser spot size to reduce
a laser beam dispersion angle in the transport line.
The spot size is continuously adjustable by shifting positions of the three lenses.
In the current design, the spot size at the transport line is 7.0 mm\footnote{
In optics, laser spot size is defined as 2$\sigma$ width of a Gaussian power distribution.}.
Components of (5) to (8) are to monitor
timings, powers, profiles, and positions of laser pulses, shot by shot.

\begin{figure}
\begin{center}
\includegraphics[width=32em]{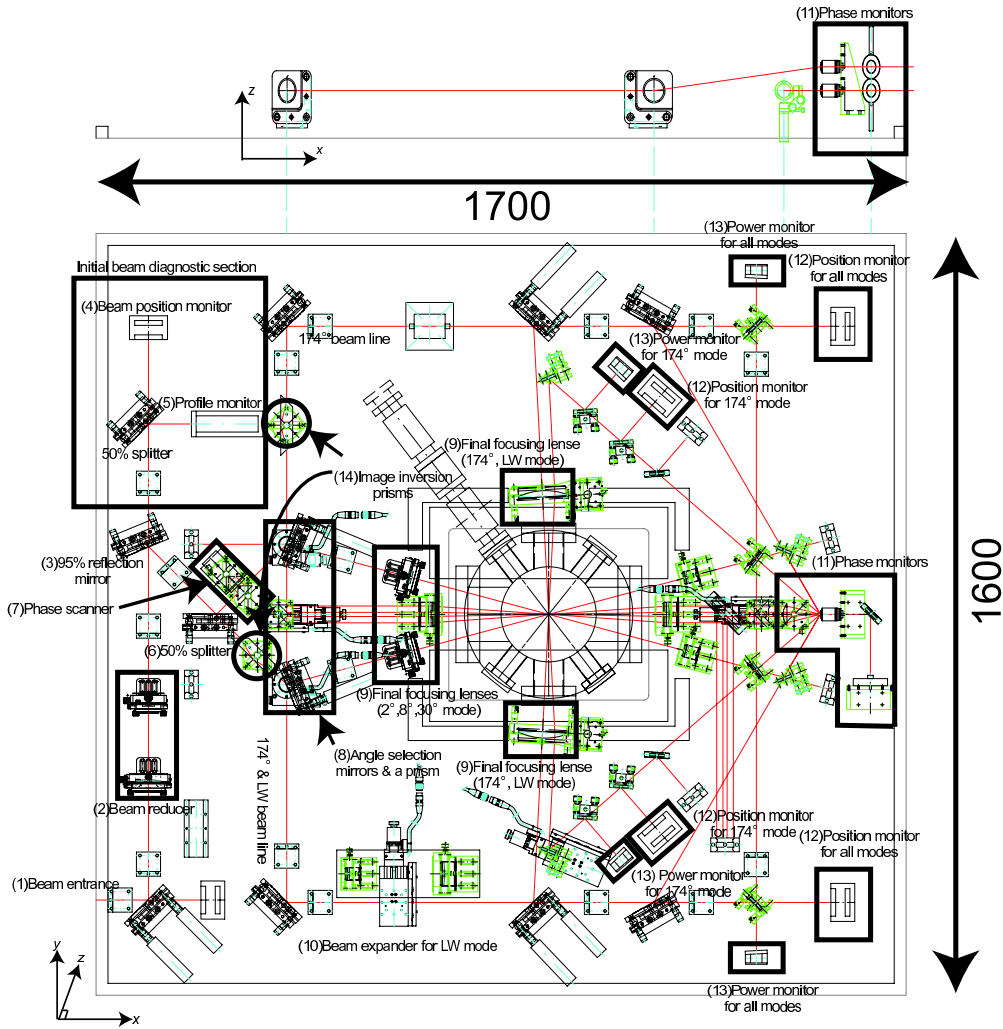}
\caption{Layout of the main optical table of the monitor.
Beam height from the table top is $z=100$ mm for all the optical components except
the phase monitor, shown in the upper figure.
}
\label{fig:optical-table}
\end{center}
\end{figure}

The laser photons are delivered to the main optical table via the transport line.
The transport line has about 15 m length, covered with metal pipes.
Several mirrors are installed in the transport line to change directions of the
laser path.
In the main optical table,
the laser beam is split into two and the split paths are intersected at the IP.
To obtain a wide observable $\sigma_y^\ast$ range from 25 nm to 6 $\mu$m,
the crossing angle of the split laser paths can be mechanically switched among
2\degree, 8\degree, 30\degree and 174\degree (See Fig.~\ref{fig:moddepth}).
The laser beams are focused at the IP to achieve high photon density.
Calculated average number of Compton photons is about 4400/bunch, using
21 $\mu$m design laser spot size (except for the 30\degree mode, which has 25.2 $\mu$m spot size
because of geometrical restrictions) at the IP.
For $\sigma_x^\ast$ measurement, a laser-wire is used instead of a Shintake monitor.
The spot size at the IP is reduced to 7.0 $\mu$m (2$\sigma$) to observe 2.8 $\mu$m $\sigma_x^\ast$.

Figure \ref{fig:optical-table} shows an optical design of the table.
The laser beam comes from bottom-left corner(1), and goes through a beam reducer(2), which reduces
the laser beam size to a half, 3.5 mm to avoid tail cuts by optical components in the table.
A 95\% reflection mirror(3)
is placed after the beam reducer. Transmittant laser beam (5\% energy) goes to a diagnostic section
which consists of a laser position monitor(PSD)(4) and a laser profile monitor(5).

Reflected beam (95\%) goes to a 50\% beamsplitter(6), which divides the laser path
to upper and lower. In the upper path just after the beam-splitter, we install a phase scanner system(7). 
Selection of the operation modes is performed
by angle selection mirrors, which are placed in both of laser paths(8).
They can select 2\degree, 8\degree, 30\degree and 174\degree crossing angle modes by rotating themselves
by stages below the mirrors.
In addition, selection of laser-wire mode needs the angle selection mirrors.
In the laser-wire mode, lower rotation mirror is set as same as 174\degree setup, and
upper mirror is set to send the laser path to the absorber.
The laser beams are focused by final focusing lenses(9).
For the laser-wire mode, a 3$\times$ beam expander(10) is inserted at the laser path 
to obtain smaller spot size at the IP. 
Planned laser spot sizes for these crossing angles are shown in Table \ref{tbl:spotsize}. 

\begin{table}
	\begin{center}
		\begin{tabular}{|c|r|r|r|}\hline
			Crossing angle & $w$ [mm] & $f$ [mm] & $w_0$ [$\mu$m] \\ \hline\hline
			2\degree, 8\degree and 174\degree 	& 3.53  & 250.1	& 21.0 \\ \hline
			30\degree 													& 3.53	& 300.1	& 25.2 \\ \hline
			laser wire 													& 10.58	& 250.0	& 7.0 \\ \hline
		\end{tabular}
	\caption{Planned laser spot sizes at the IP. $w$, $f$ and $w_0$ stand for
					spot sizes at the focal lenses, focal lengths of the lenses, and spot sizes at the IP, respectively.
$M^2 = 1.75$ (measured value) is used to calculate spot sizes.}
	\label{tbl:spotsize}
	\end{center}
\end{table}

The right half of the table is mainly for diagnostics. A couple of Phase monitors is shown as (11).
Each of them consists of an objective lens with an image sensor.
Delivered paths of laser beams to the objective lenses are different for each crossing angle.
Other monitors in the right side are position sensitive detector (PSD)s(12)
and photodiode (PD)s(13) for position feedback, intensity jitter correction
and accurate alignment.
We implement the optical layout so that a couple of PSDs and PDs can cover all crossing angles.
In 174\degree mode other PSDs and PDs are installed for better stabilization.
Image inversion prisms(14) are installed to arrange the image directions of the laser beam at the IP
in parallel. This arrangement significantly suppress contrast degradation of the fringe at the IP
caused by laser beam position fluctuations.

\begin{figure}
\begin{center}
\includegraphics[width=25em]{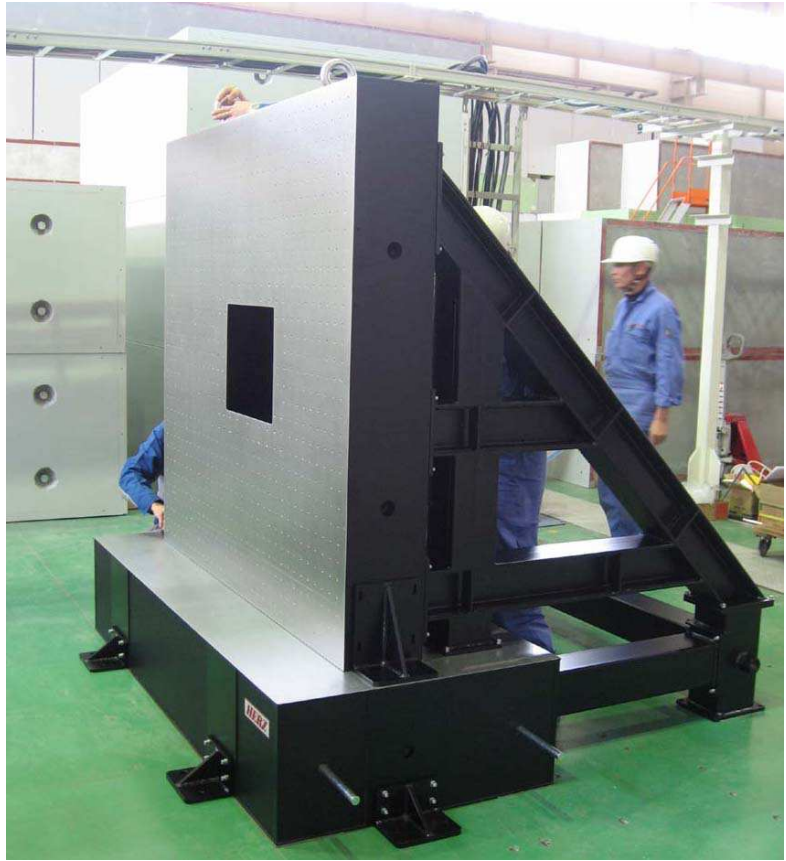}
\caption{The main optical table with a support frame. Optical parts and a vacuum chamber
will be attached on the vertical plane.
}
\label{fig:table-pic}
\end{center}
\end{figure}

To suppress the vibration of the table with respect to the electron beam, a rigid support frame is fabricated.
Figure \ref{fig:table-pic} shows a picture of the table with the support frame.
The main table is made of a 250 mm thickness steel honeycomb core.
Total weight is about 2000 kg, including weight of the main table (700 kg).
The table is fixed to the ground independent of the final focusing magnets.
Since the table of the final focusing magnets is also attached firmly to the ground,
both the Shintake monitor and the focusing magnets are expected to move together with the ground.
10 nm level stability of the relative motion between the Shintake monitor and the magnets
is targetted.

\subsection{Gamma Detector}

\begin{figure}
\begin{center}
\includegraphics[width=20em]{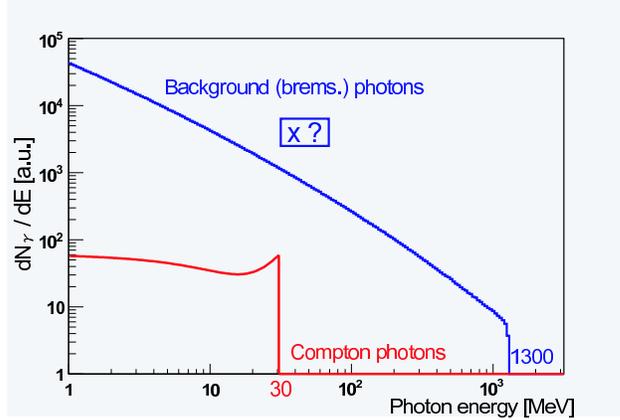}
\caption{Energy spectrum of Compton signal and beam background.
Since ratio of signal and background is unknown,
vertical axis is arbitrary scaled for each spectrum.}
\label{fig:gammaenergy}
\end{center}
\end{figure}

The gamma detector is installed on the beam line after the IP, besides the beam dump.
Figure \ref{fig:gammaenergy} shows energy spectra of Compton photons (singal) and
background. Drawn background is from the beam pipe scattering with beam halo electrons.
Compton photons have maximum energy of around 30 MeV, and background photons have
broader energy range up to 1.3 GeV.
Since the amount of background is expected not to be negligible,
we need to suppress or subtract the background.

\begin{figure}
	\begin{center}
	\begin{minipage}[t]{.47\textwidth}
		\includegraphics[width=15em]{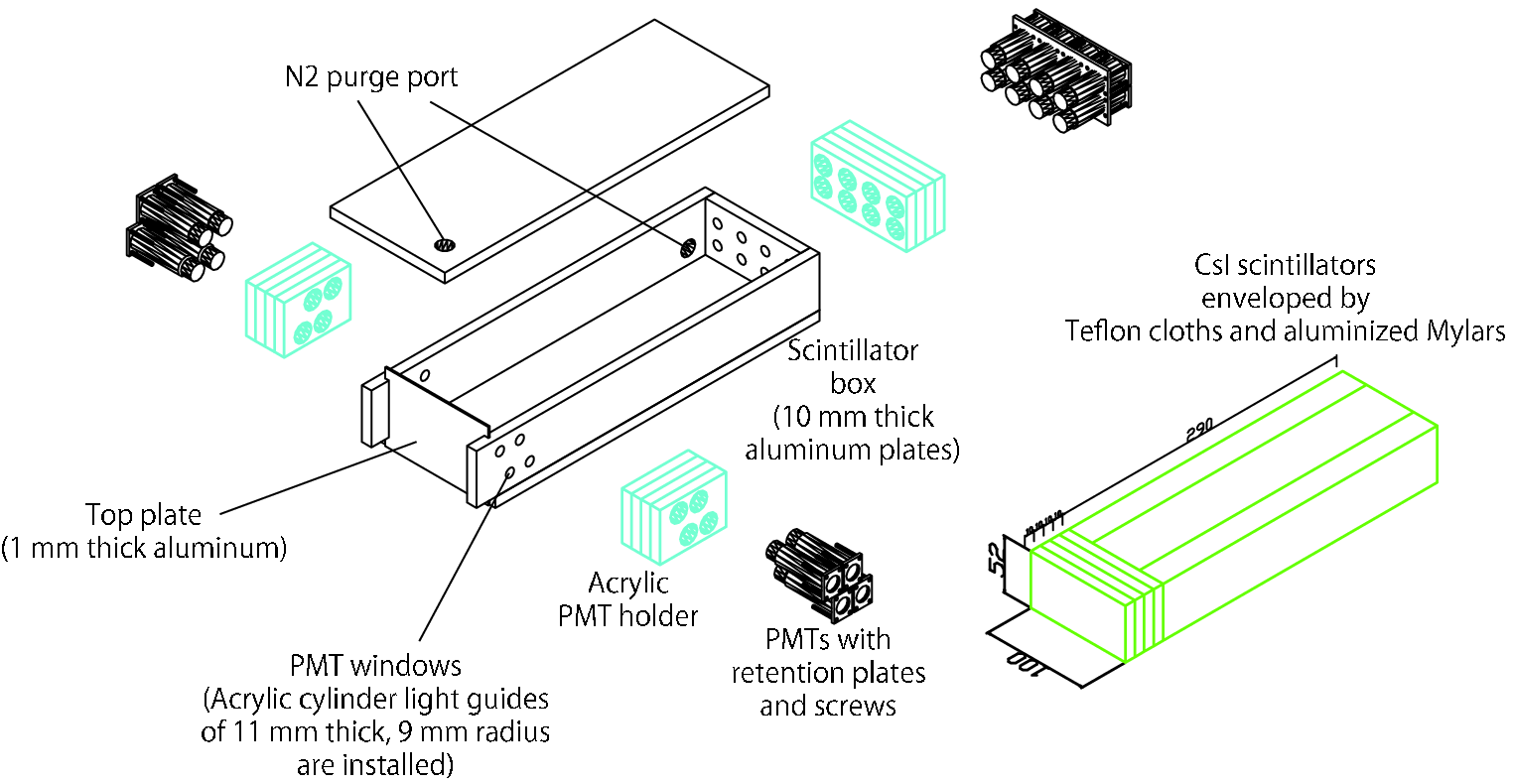}
	\end{minipage}
	\hfill
	\begin{minipage}[t]{.47\textwidth}
		\includegraphics[width=15em]{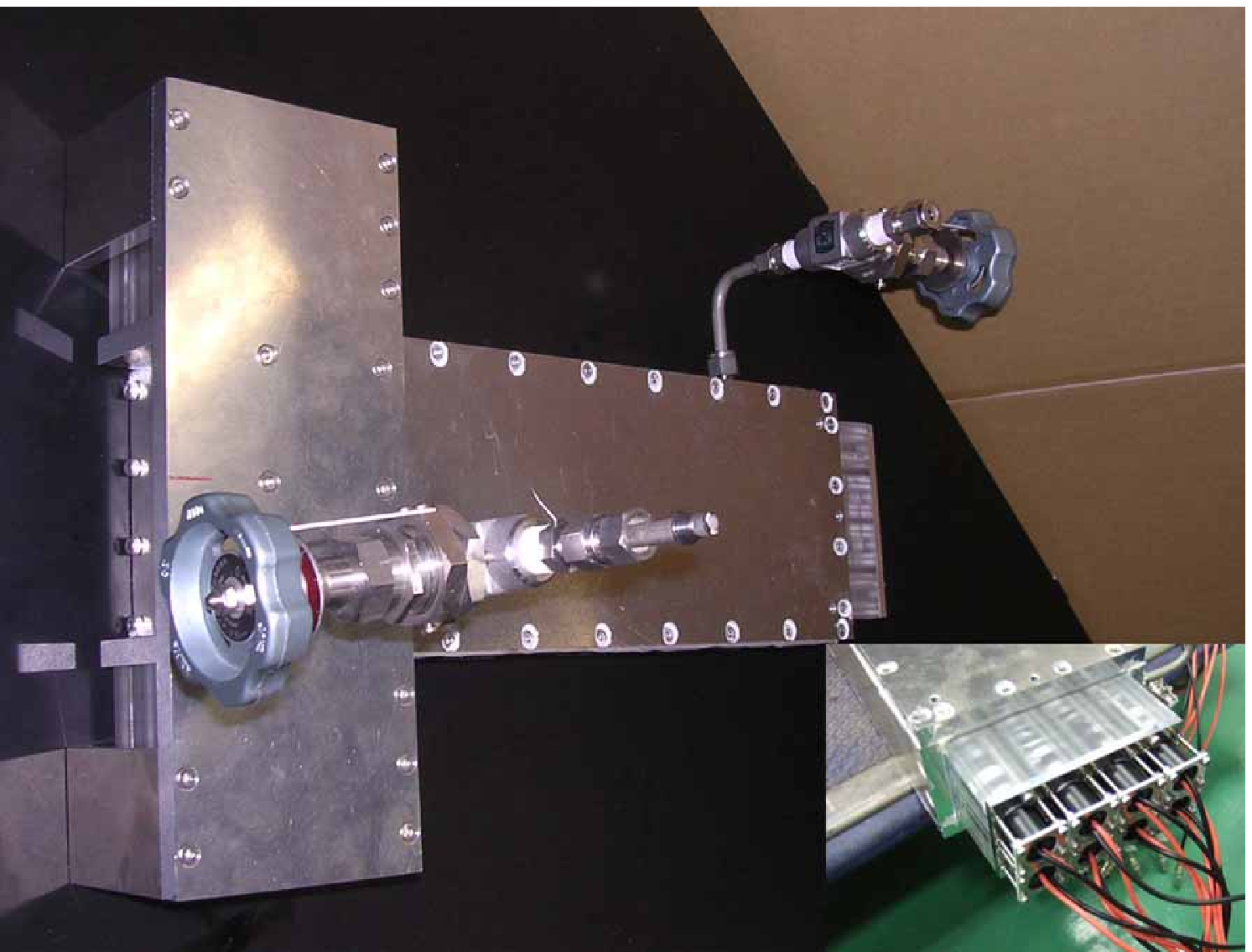}
	\end{minipage}
	\caption{Components drawings and a picture of the gamma detector.}
	\label{fig:detector}
	\end{center}
\end{figure}

Figure \ref{fig:detector} shows a drawing and a picture of the detector.
The detector consists of 4-layer forward scintillators and 3 pieces of bulk scintillators.
Scintillators are made by thallium-doped cesium iodide (CsI(Tl) : 1.86 cm $X_0$) crystals.
Total scintillator size is 100 mm width, 50 mm height, and 330 mm (17.7 $X_0$) along the beam axis,
enough volume for $<$ 1.3 GeV photons.
Each scintillator piece is optically separated
by wrappings of a 200 $\mu$m-thick Teflon sheet and a 50 $\mu$m-thick alminized Mylar.
The forward scintillators are 10 mm thick (0.54 $X_0$) along the beam axis, 
equipped with photomultiplier tubes (PMT) on each side.
The bulk scintillators have 290 mm (15.6 $X_0$) length, 25 mm (side) or 50 mm (center) width and 50 mm height.
Two (side) or four (center) PMTs are attached at the end side of each scintillator.
Since CsI(Tl) crystal is slightly hygroscopic, we made a semi-airtight container 
with sealing by silicon glue and sheet, to close up all the scintillators together. 
The container is also equipped with a couple of gas inducers,
purging by dry nitrogen periodically.
The equipped PMTs are Hamamatsu R7400U, which have 8 mm active diameter, operated with positive HV power supply.
The PMTs are mechanically attached to the container with support structures.
Acrylic cylinder light guides are installed and glued to the container, and
PMTs are touched to the light guides. No optical cements are used
to enable easy replacement of PMTs.
Signals from PMTs are sent to a charge-sensing ADC.
The gate of the ADC is planned to be derived from BPM signal,
which has almost no jitter.

Detailed description and performance estimation including calibrations and
beam tests are described elsewhere\cite{detector}.

\section{Laser Position Alignment and Stabilization}
\label{sec:laserpos}

Alignment and position stabilization of the laser spots at the IP are critical issues
for the Shintake monitor.
Since the laser spot position errors degrade beam size measurements,
we need to minimize and/or compensate the position errors.

\subsection{Effects of Displaced Laser Beams}

\begin{figure}
\begin{center}
\includegraphics[width=30em]{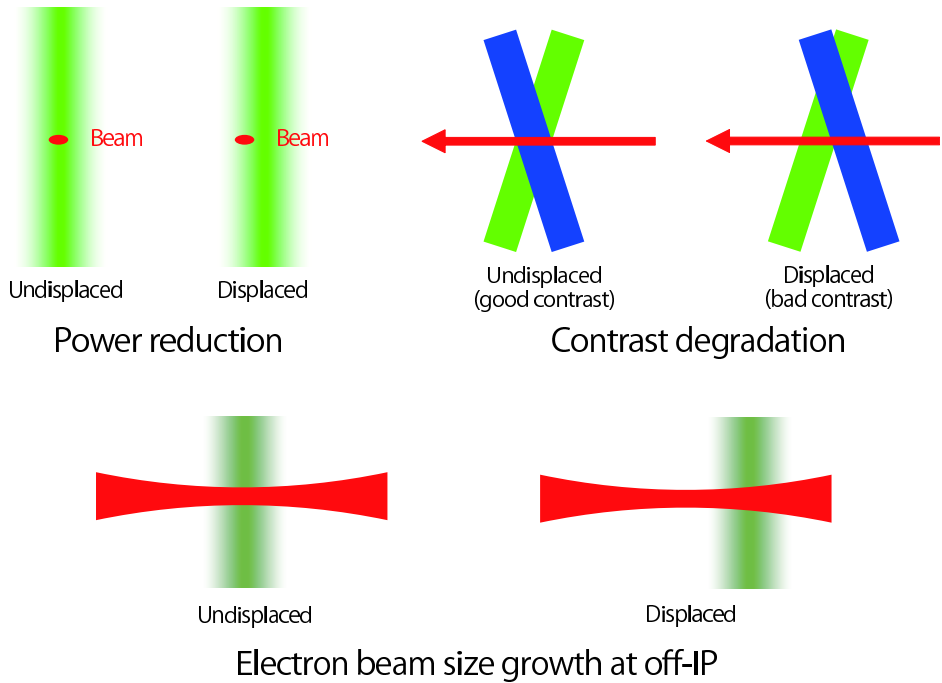}
\caption{A schematic view of the three effects by the displaced laser beams.}
\label{fig:table:poserror}
\end{center}
\end{figure}

Firstly we overview effects of displaced laser beams, caused by misalignment and instability
of the laser beam.
The effects of the displaced laser beams are categorized into three error factors:
power reduction, contrast degradation and electron beam size growth.
Figure \ref{fig:table:poserror} shows a schematic view of these three effects.
Both short-term position instability and long-term position displacement (or misalignment)
should be considered to understand these effects.

\subsubsection{Power Reduction}

A displacement of the laser beam causes a deviation of photon density at the IP.
The photon density at the IP with a displacement of laser beam position on the $x$-$y$ plane
is described as,
\begin{equation}
	P(\delta_x) = P_0\exp\left(-\frac{\delta_x^2}{2\sigma^2}\right),
\label{eqn:posfluc-power}
\end{equation}
where $\delta_x$ is the displacement, $\sigma$ is the laser spot size,
and $P_0$ is the photon density at no displacement.

For this error factor, 
we only need to consider the short-term position instability
since the long-term position displacement only causes a constant power reduction
and then it does not give significant effects to the modulation measurements.
However, when average laser spot position is displaced, power fluctuation due to the
position fluctuation is enhanced, since the $P(\delta_x)$ has flat distribution
around the $\delta_x = 0$ and non-flat at the larger $\delta_x$.
Consequently, we should also care the long-term position displacement
if non-negligible position fluctuation is inevitable.

\subsubsection{Contrast Degradation}

Contrast degradation is induced by the relative displacement of two crossing laser beams.
The contrast is affected by both short-term and long-term position displacements.

Contrast degradation is caused by power imbalance of the two laser spots, given by\cite{shintake1-2},
\begin{equation}
	M_\gamma = \frac{(A_1+A_2)^2-(A_1^2+A_2^2)}{A_1^2+A_2^2} = \frac{2A_1A_2}{A_1^2+A_2^2}
		= \frac{2\sqrt{P_1P_2}}{P_1+P_2}
							\equiv \frac{2\sqrt{P_A}}{1+P_A}
\label{eqn:posfluc-contrastdeg1}
\end{equation}
where $A_1$ and $A_2$ are amplitudes of the 2 light paths, $P_1$ and $P_2$ are powers of them,
which are square roots of $A_1$ and $A_2$.
$P_A$ is defined as $P_2/P_1$, which shows the power imbalance.

Response of the spot displacement to the power imbalance is different between
displacement along the $x$-$y$ plane and along the $z$ axis.

For the displacement along the $x$-$y$ plane, the contrast degradation is given by,
\begin{equation}
	M_{\gamma,\delta_{x}} = \frac{2\exp\left(-\frac{\delta_{x1}^2}{4\sigma^2}\right)}
								{1+\exp\left(-\frac{\delta_{x1}^2}{2\sigma^2}\right)},
\end{equation}
where $\delta_x$ is the displacement and $\sigma$ is the laser spot size.
In this calculation the electron beam is assumed to pass through the peak of the one of the laser spot
(the worst case for the contrast degradation).

For the displacement along the $z$ axis,
\begin{equation}
	M_{\gamma,\delta_z} = \exp\left(-\frac{\delta_z^2}{8\sigma^2}\right).
\end{equation}
where $\delta_z$ is the displacement.
The contrast is usually much sensitive to displacements along the $z$ axis than along the $x$-$y$ plane.

\subsubsection{Electron Beam Size Growth at the Off-IP}

To achieve extremely small electron beam size at the IP, the electron beam is strongly focused
at the IP, with a wide dispersion angle. Because of this wide dispersion angle, waist length of
the electron beam around the IP is very short.
If the laser spot position fluctuates along the $z$ axis, the effective beam size is enhanced.

Vertical electron beam size $\sigma_y$ depends on the position along the electron beam axis($z$) as,
\begin{equation}
	\sigma_y(z) = \sqrt{\beta_y^\ast\epsilon_y\left(1+\frac{z^2}{\beta_y^{\ast2}}\right)}
							= \sigma_y(0)\sqrt{1+\frac{z^2}{\beta_y^{\ast2}}}
\label{eqn:table:bsgrowth}
\end{equation}
where $\beta_y^\ast$ is the beta function at the focal point which is 100 $\mu$m in ATF2,
and $\epsilon_y$ is the vertical emittance.

\subsection{Alignment}

In this subsection we discuss about the laser beam position alignment at the IP.
Position alignment along the $x$-$y$ plane is performed by beam scan
and alignment along the $z$ axis is performed by slit scan.

\subsubsection{Beam Scan}
\label{sec:beamscan}

\begin{figure}
\begin{center}
\includegraphics[width=20em]{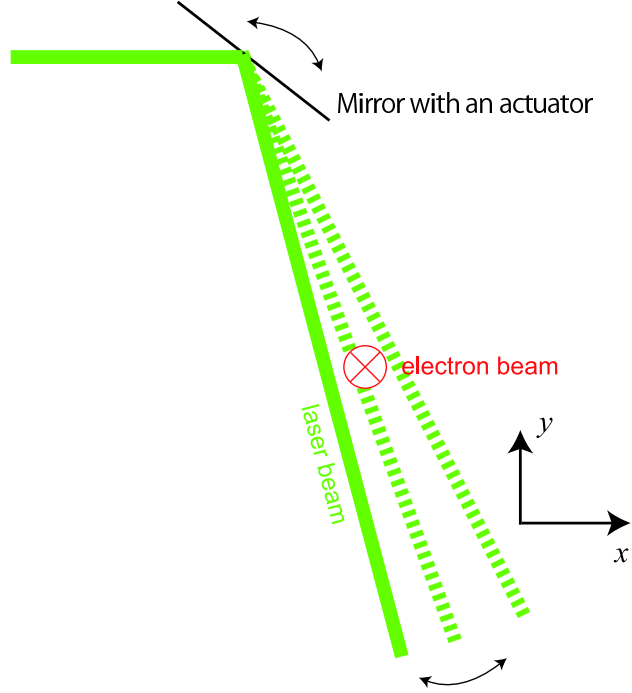}
\caption{A schematic of the beam scan along the $x$-$y$ plane.
The electron beam is scanned by the laser beam with an actuator at the final focus lens.
Compton scattered photons are monitored by the gamma detector to obtain the alignment position,
which is the position where the maximum number of photons is obtained.
}
\label{fig:beamscan-schematic}
\end{center}
\end{figure}

For an alignment along the $x$-$y$ plane, we use the electron beam itself to
cross two laser beams just on the electron beam line.
Figure \ref{fig:beamscan-schematic} shows a schematic of the beam scan.
Beam position at the IP can be shifted by an actuator or a stage
at the forward mirror, to scan the electron beam.
When the laser beam is just at the electron beam position, 
Compton scattered photons are emitted and they can be monitored
by the gamma detector of our monitor.
We set the alignment position where the Compton signal strength is the maximum.

For the beam scan, only one laser path is introduced to the IP at a time, because
we must avoid forming the interference fringe during the beam scan.
The other beam path is sent to the absorber by the rotation stage.

We performed a toy Monte Carlo simulation to estimate alignment accuracy
of the beam scan. With estimated error factors
(4.2\% power, 2.5 $\mu$m position, and 8.3\% background jitter),
0.6 $\mu$m alignment accuracy can be obtained\cite{dron}.

\subsubsection{Slit Scan}

\begin{figure}
\begin{center}
\includegraphics[width=30em]{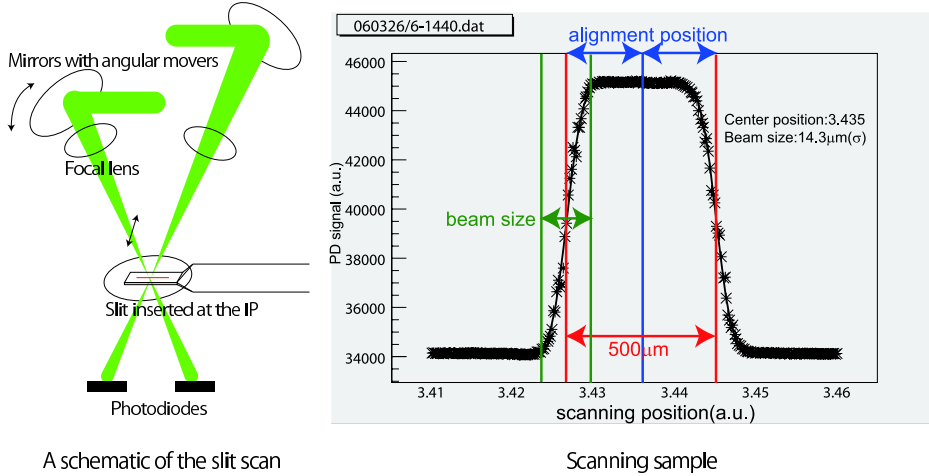}
\caption{Sample data of slit scan by low power test laser.}
\label{fig:slitscan-sample}
\end{center}
\end{figure}

The left figure of Fig.~\ref{fig:slitscan-sample} shows a schematic of the alignment along the $z$ axis
called a ``slit scan".
A slit made by stainless steel is inserted at the IP.
Mirrors equipped with actuators can steer the laser beam across the slit,
and photodiodes after the IP monitor
the arrived light intensity which passed through the slit.
The right figure of Fig.~\ref{fig:slitscan-sample} shows the detected photodiode intensity
during the slit scan. The slit width is 500 $\mu$m, which is much wider than the laser spot size.
Obtained graph can be expressed as,
\begin{eqnarray}
	P(x) &=& C\left\{\mathrm{erf}\left(\frac{x-\mu+w/2}{\sigma}\right)
				 - \mathrm{erf}\left(\frac{x-\mu-w/2}{\sigma}\right)\right\} \\
	\mathrm{erf}(x) &\equiv& \frac{1}{\sqrt{2\pi}}\int^x_0 e^{-t^2/2}dt
	\label{eqn:slit-erf}
\end{eqnarray}
where $x$ is the laser position, $C$ is amplitude,
$\mu$ is the center position of the slit, $w$ is the width of the slit,
and $\sigma$ is the laser spot size (1$\sigma$). $\mathrm{erf}(x)$ is an integral of Gaussian function,
called ``error function". The line in Fig.~\ref{fig:slitscan-sample} shows the fitting result
by $P(x)$. We can adopt fitted $\mu$ for the alignment target.

Positions of the laser spots are monitored by PSDs.
Accuracy of the alignment is limited by linearity of the PSDs, which is $<$ 9.1 $\mu$m (measured value).
Position uncertainty caused by this linearity is $<$ 1.7 $\mu$m by optics calculations.

To align the electron beam waist to the center of the laser spots,
electron beam waist position is adjusted
by controlling strength of magnetic field of the final quadrupole.
Required position accuracy of the beam waist is about 20 $\mu$m,
which is achievable by this method.

Slit scan is also used for the laser spot size measurement at the IP.
The spot size is obtained by fitting data around the edge of the slit by an error function
(\ref{eqn:slit-erf}).
The measurement is fluctuated by the laser power jitter and the position jitter.
We performed a toy Monte-Carlo simulation with estimated error factors
(1\% laser power, 2.5 $\mu$m position jitter),
and obtained that accuracy of spot sizes is 0.9 $\mu$m.

\subsection{Stabilization and Correction}

As transient position displacements of the laser spots,
angular jitter of the laser and slow drift (by temperature etc.) should be considered.

\subsubsection{Correction of the Laser Angular Jitter}

An intrinsic angular jitter of the pulsed laser is the largest source which
causes pulse-to-pulse laser position fluctuations.
Measured value of the angular jitter is about 10 $\mu$rad., while it may vary
by environmental conditions.
This jitter cannot be actively stabilized, but it can be monitored by PSDs.
Since the position displacement caused by the angular jitter is enhanced by
the long transport line, the jitter can be monitored precisely by PSDs.
Using the measured RMS pulse-to-pulse resolution of our PSDs 9.1 $\mu$m,
1.0 $\mu$rad.~resolution is achievable.

The measured angular jitter is used to correct the power fluctuation 
caused by the angular jitter.
In case of 10 $\mu$rad.~angular jitter, the power fluctuation is 3.9\% without correction.
It can be corrected within 1.4\% accuracy using the measured angular jitter
(toy Monte-Carlo estimation again).
0.6 $\mu$m alignment accuracy along the $x$-$y$ plane (see Section \ref{sec:beamscan}) is used
for the estimation.

\begin{figure}
\begin{center}
\includegraphics[width=30em]{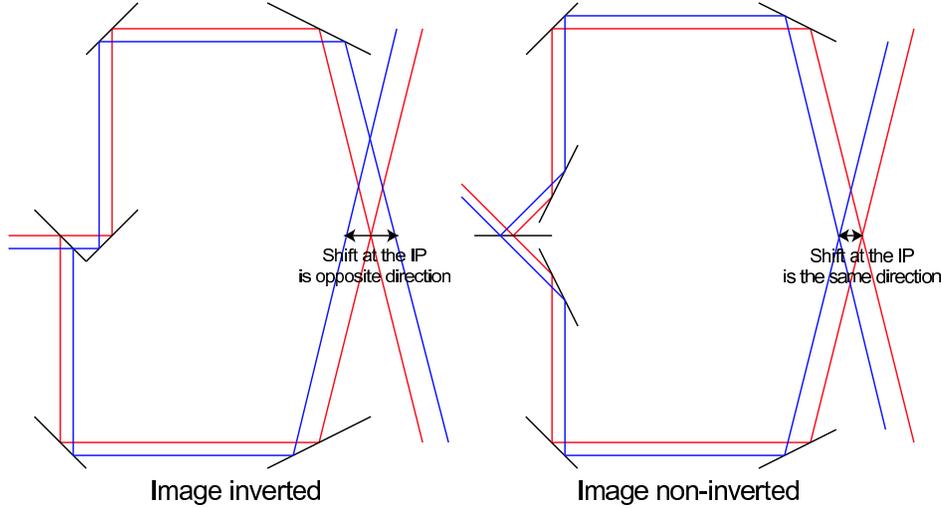}
\caption{A schematic of the image inversion. If the beam line is image inverted (Left figure),
the position displacement at the laser or forward optics causes displacements of 
the opposite direction at the IP. If the beam line is non-inverted (Right figure),
it causes displacements of the same direction at the IP.
}
\label{fig:imagedirection}
\end{center}
\end{figure}

For the contrast degradation, selecting proper ``image direction" enables
to cancel the displacement between the two laser paths.
Figure \ref{fig:imagedirection} shows a schematic of the image inversion.
The blue and red laser beams are positionally displaced, which causes the position 
deviation at the IP. In the image-inverted (left) figure, the direction of the 
displacement at the IP is the opposite direction to each other, while
in the image non-inverted (right) figure, the direction is the same.

The figure shows the case of position displacement,
but the response of the angular jitter is the same as the position
displacement.
To suppress the contrast degradation by the laser angular jitter,
the image direction is designed to be the same for the incoming beam lines
in our optical table, using Dove prisms to flip the image direction.

With the proper image direction, contrast degradation caused by the angular jitter
can be suppressed to negligible level.

For the off-IP beam size growth, the position jitter is negligible,
because 20 $\mu$m position displacement (1 order of magnitude larger than the
displacement caused by measured angular jitter) only causes 2\% beam size growth.

\subsubsection{Stabilization of Slow Drift}

\begin{figure}
\begin{center}
\includegraphics[width=30em]{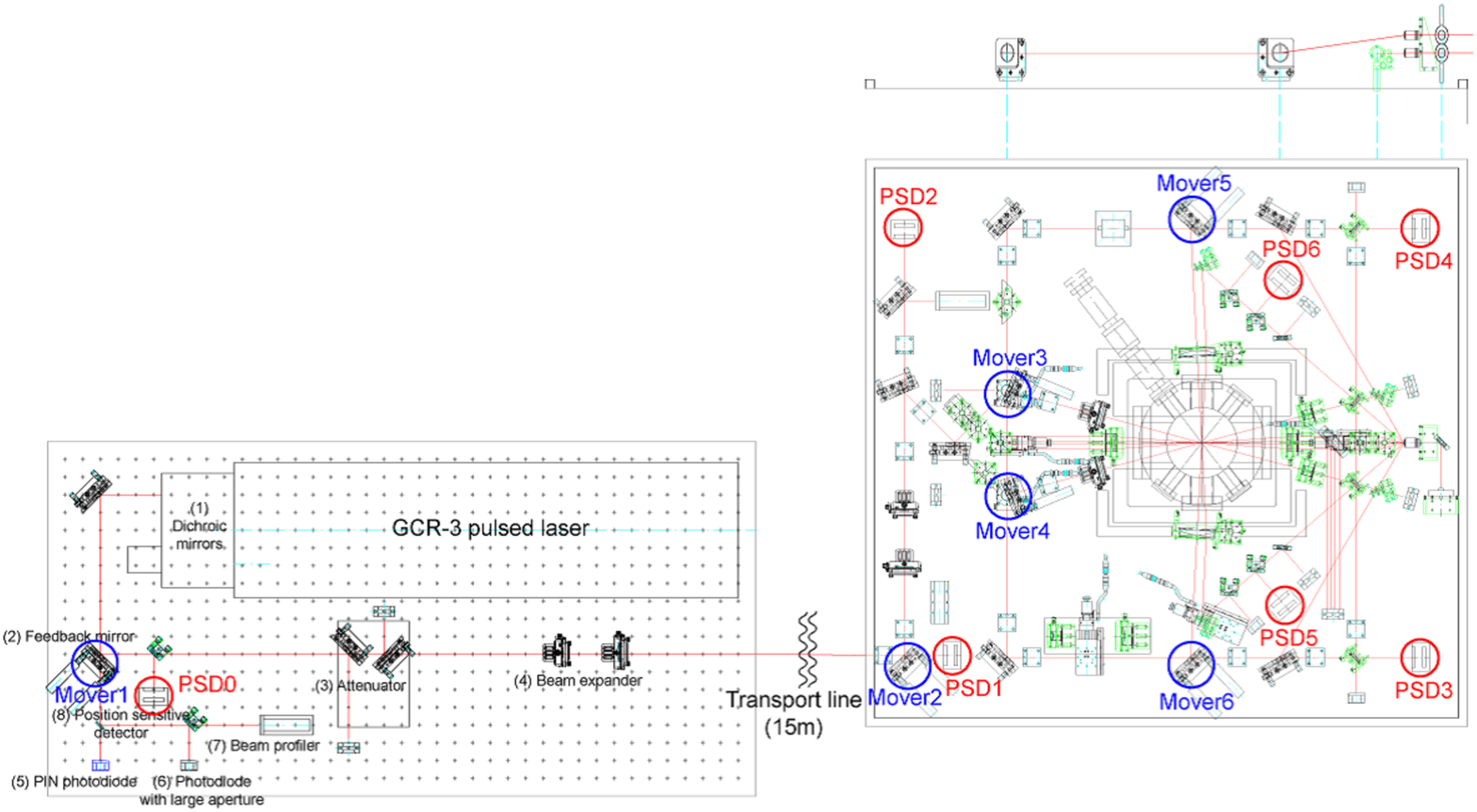}
\caption{Position of the actuators and the position sensitive detectors.}
\label{fig:mover-psd-pos}
\end{center}
\end{figure}

The slow laser beam position drift (timescale larger than minute) can be induced
by plenty of error sources, laser drift, physical shift of the laser transport line,
angular shift of the optical components on the optical table, etc.
Since the measurement time of the Shintake monitor is 1 minute,
the slow drift causes only an alignment error.
It can be canceled by active stabilization using USDs and mirror actuators.

The installed location of the PSDs are shown in Fig.~\ref{fig:mover-psd-pos}.
A geometrical laser path (which is a drift space without any mirrors or lenses)
is defined by 4 variables, center position (x,y) on a certain plane
and forward angle (azimuth and elevation).
To stabilize a laser path, we need 2 PSDs and 2 actuators for both position and angle. 

The drift from the origin of the laser and the transport line (outside-origin drift)
has larger amplitude than the drift from the origin on the optical table (inside-origin)
because laser has a lot of instability and
the environmental condition is worse for the transport line.
The outside-origin drift is canceled by locally PSD1 and PSD2 with Mover1 and Mover2,
to prevent the outside-origin drift from affecting the optics on the table.
PSD 3 - 6 with Mover 3 - 6 are mainly for the alignment, but the inside-origin drift
can also be canceled using these PSDs and movers.

Because of the large angular jitter of the laser, we need to average $\sim$ 100 pulses
for the active stabilization with high accuracy (better than alignment accuracy).
Therefore, stabilization of $<$ 1 minute is not feasible.
We expect that the drift within the measurement time is smaller than the
alignment accuracy.


\section{Phase Control}
\label{sec:fringe}

Fluctuations of the relative phase of the two laser beams at the IP
cause fluctuations of the fringe pattern.
To suppress the phase fluctuation, an active phase control system
is implemented in our Shintake monitor.
In addition, phase scanning to obtain modulation spectrum is also
performed using the phase control system.
The phase control system consists of phase monitors and a phase
mover with a active feedback software.

\subsection{Phase Monitor}

\begin{figure}
\begin{center}
\includegraphics[width=30em]{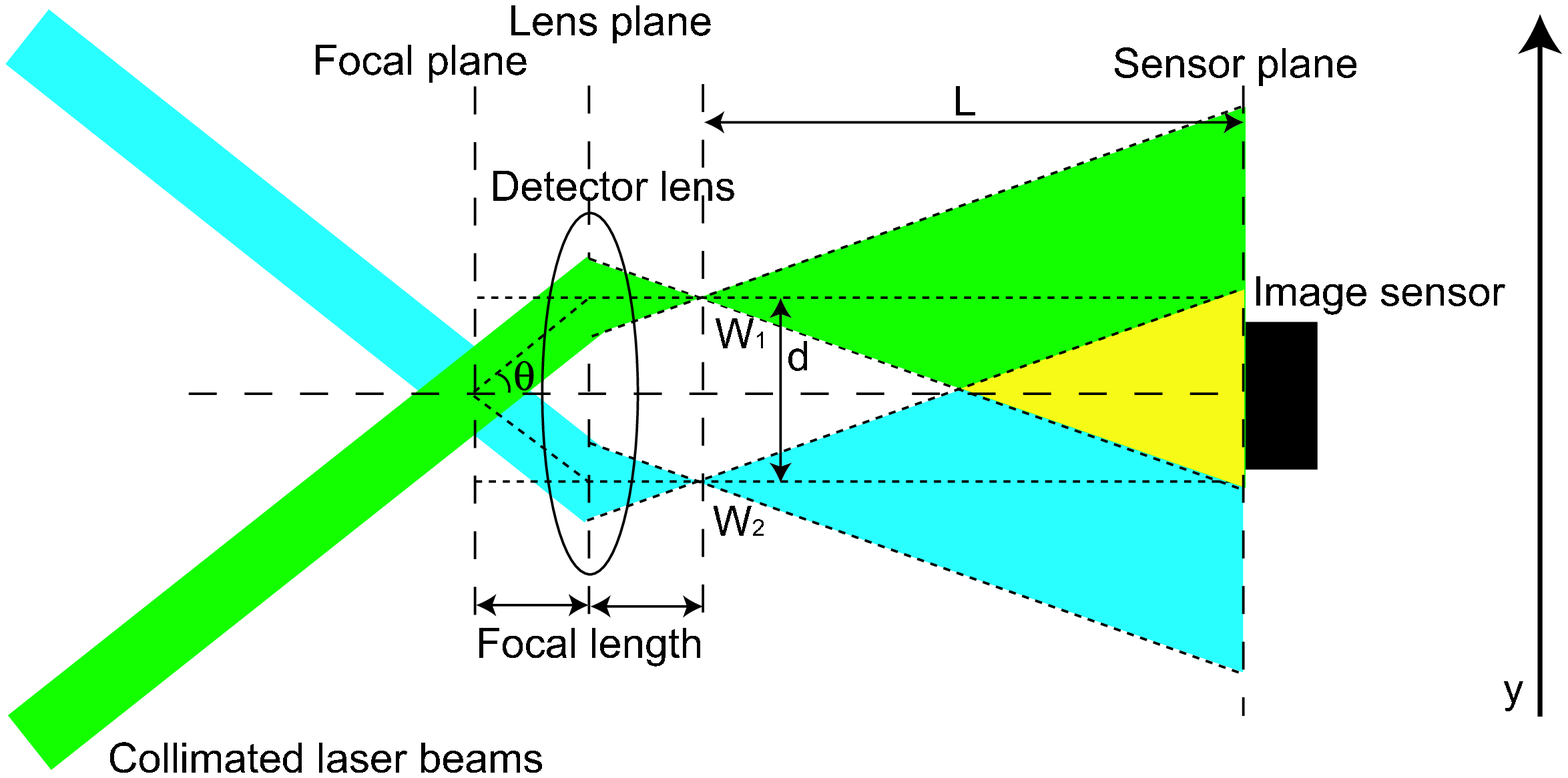}
\caption{Schematic of fringe magnification by a lens. Approximation of geometrical optics
is applied in this figure.}
\label{fig:lensmag}
\end{center}
\end{figure}

A schematic drawing of a phase monitor is shown in Fig.~\ref{fig:lensmag}.
The phase monitor consists of a lens for fringe magnification and an image sensor for
fringe pattern acquisition.

Split laser beams are collimated and introduced to the detector lens as
the center of the laser beam goes through just the focal point of the lens.
Since light from the focal point goes parallel to the lens axis
after passing through the lens,
the center of the laser beam goes along the lens axis.
Surrounding light of the laser beam is focused on the opposite focal plane and then diverged
as shown. Interference fringe is formed in the overlapped area (painted yellow).
Obtained fringe phase at the image sensor corresponds to a phase difference of the two laser
beams at the focal plane.

To accept large crossing angles, focusing power of the detector lens must be very large.
We use an objective lens whose focal length is 2 mm (100$\times$ multiplication),
NA (numerical aperture) = 0.95.
Crossing angles up to 144\degree can be monitored via the objective lens.
With the objective lens, we can obtain clear interference fringes,
which can be observed by the image sensors.

A CMOS linear image sensor is used for the phase acquisition.
It has 1024 pixels in 7.8 $\mu$m pixel pitch, and its readout frequency is 187 kHz / pixel.
The pixel data are sent to a 200 kHz 12 bit VME ADC.
Data from two image sensors can be read interleaved in 10 Hz
repetition rate (corresponding to every laser pulse).

To extract phase information from the waveform data, we use Fourier transformation method.
Fourier transform is defined as,
\begin{equation}
	g(\omega) = \int^{+\infty}_{-\infty}f(t)e^{-i\omega{}t}dt
\end{equation}
where $f(t)$ is an original function and $\omega$ is an angular velocity, which is the base of
the transformation. Physically, $|g(\omega)|$ represents fraction of power at the angular velocity
of $\omega$, and $\arg{}g(\omega) = \tan^{-1}\left(\mathrm{Im}\ g(\omega)/\mathrm{Re}\ g(\omega)\right)$
represents phase at $\omega$.

A fringe pattern makes a sine curve at the image sensor.
A sine function is converted to a delta function which has a peak at the sine frequency by Fourier transform.
We can get the phase of the sine function by calculating $\arg{}g(\omega)$ at the delta-peak frequency.

Fourier transform assumes that $f(t)$ is perfectly smooth from $-\infty$ to $+\infty$,
while in fact the waveform is discrete and restricted in a finite region.
In real analysis, we modify the transform as
\begin{equation}
	g(j) = \sum^{N/2-1}_{k=-N/2}f_k\exp\left(-\frac{2\pi{}ijk}{N}\right)\qquad(0<j<N).
\label{eqn:fringe:dftmodified}
\end{equation}
where $f_k$ is a point of discrete waveform.

This modification is slightly different from ordinal Discrete Fourier Transform (DFT).
First, the ordinal DFT calculates only integer $j$, though fractional $j$ is also used for
our analysis to obtain better resolution of the peak frequency.
Second, summing range of $k$ is shifted to ($-N/2$,$N/2-1$), compared to
the ordinal DFT (0,$N-1$). This modification suppresses a phase jitter caused by a position
error of the peak frequency.

\begin{figure}
\begin{center}
\includegraphics[width=30em]{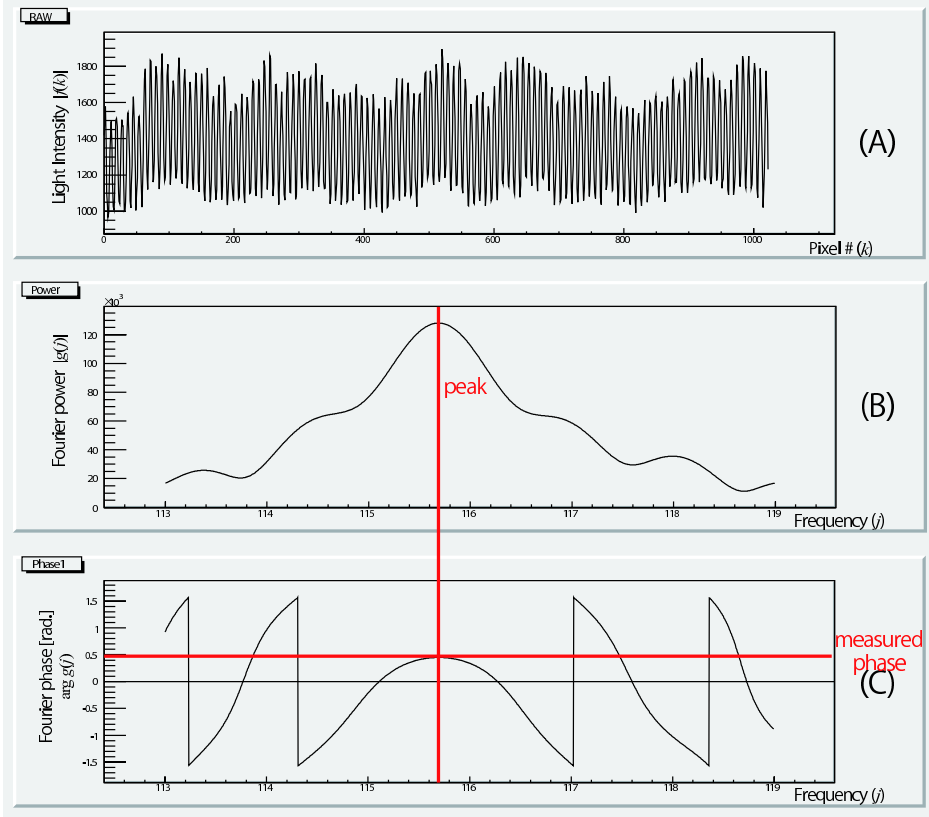}
\caption{A detection sample of Fourier transform.
(A): a raw waveform captured by the image sensor.
(B): Fourier power spectrum $|g(j)|$ around the peak.
We can observe a peak of power spectrum at $f\sim116$.
(C): the phase spectrum, $\arg{}g(j)$ of (\ref{eqn:fringe:dftmodified}).
The red line shows the peak position of the Fourier power spectrum.
}
\label{fig:fourier-sample}
\end{center}
\end{figure}

Figure \ref{fig:fourier-sample} shows sample waveforms of a phase acquisition.
In Plot (A) (raw waveform), a clear fringe pattern is observed.
Plot (B) shows a Fourier power spectrum. The fringe pattern in Plot (A) corresponds
to the peak around $j=115.7$ in Plot (B) (zoomed around the peak).
Plot (C) is a Fourier phase spectrum.
Measured phase in this sample is shown as the horizontal line of the plot,
phase value at the peak of the power spectrum.
This measured phase is used for reference data of the phase control.


\subsection{Phase Control}

To adjust and sweep the fringe phase, we install a phase mover on one of the split
laser paths. It consists of a piezoelectric stage and prisms to form an optical delay line
of variable length. The fringe phase at the IP depends on the difference of the two split laser
path length, then the phase can be controlled by adjusting one of the path length using the stage.

\begin{figure}
\begin{center}
\includegraphics[width=30em]{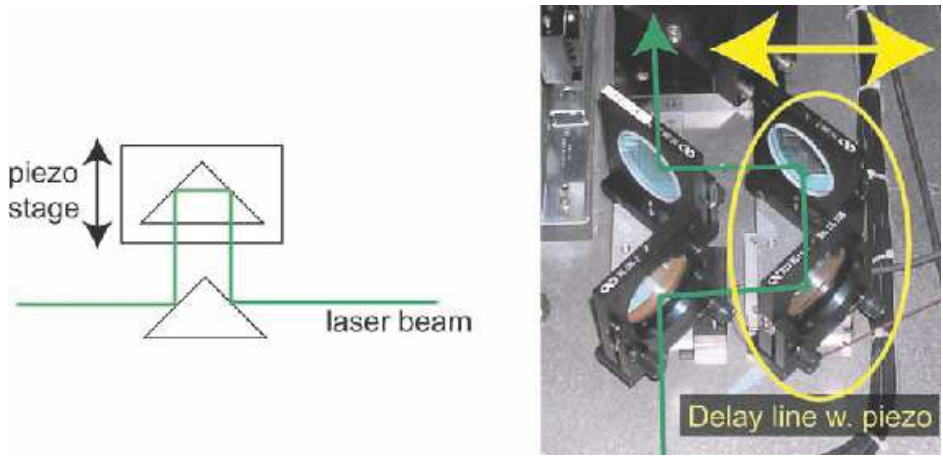}
\caption{A schematic figure of the variable optical delay line (left) and a picture of its test setup (right).
The test setup uses mirrors instead of prisms.
}
\label{fig:delayline}
\end{center}
\end{figure}

Figure \ref{fig:delayline} shows a schematic of the optical delay line.
The laser path is reflected at the orthogonal planes of the bottom prism (with high reflection coatings),
and folded by the top prism (total internal reflections).
The top prism is on a piezoelectric stage, which has 0.2 nm closed-loop resolution.
The path length can be tuned using the stage at 0.4 nm resolution.
Response time of the stage is fast enough for $<$ 10 Hz phase control system.

The stage accepts a position input by an analog voltage.
The position input signal is controlled by a VME 16 bit D/A board with a phase control software.
In the phase control software, the image sensor is triggered soon after laser pulse, and
the fringe waveform is acquired via VME ADC board. Phase calculation is performed
using the waveform by Fourier analysis.
After the calculation, the position input signal is set to cancel the deviation of the monitored phase
to the target value.

\subsection{Monitoring Location}

\begin{figure}
\begin{center}
\includegraphics[width=30em]{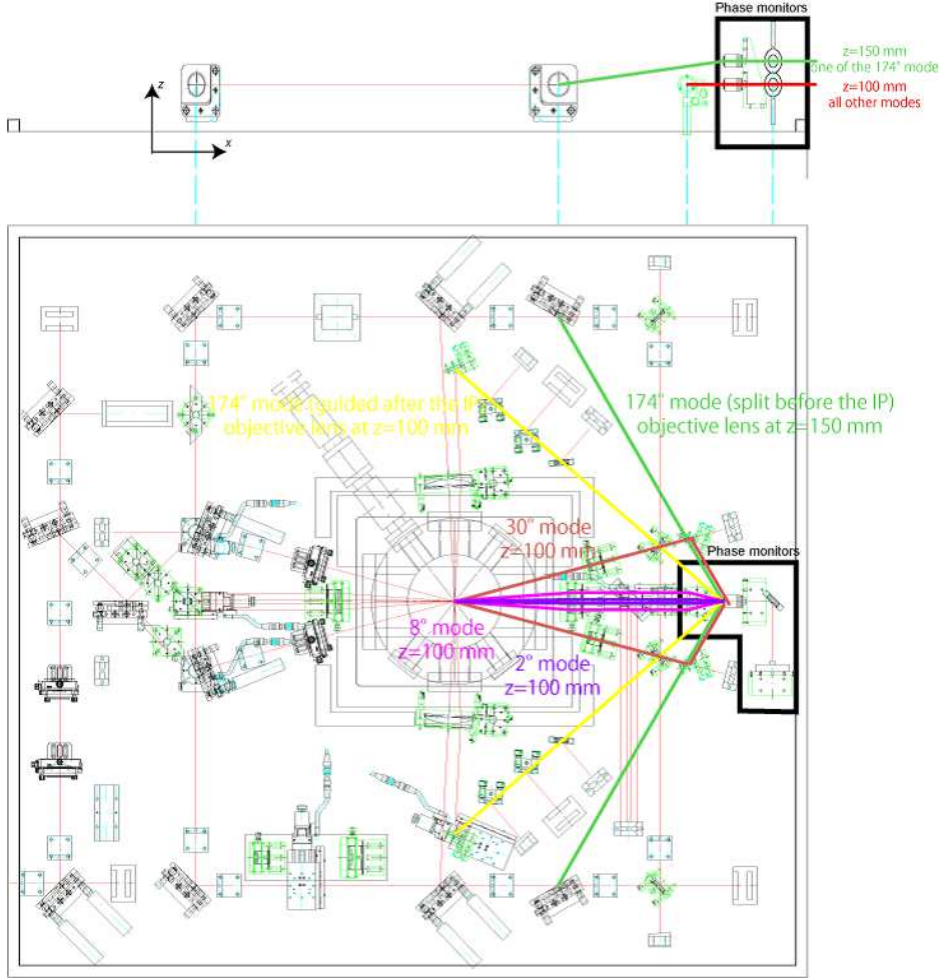}
\caption{Locations and laser paths of the phase monitors.
The phase monitors are installed at $z = 100$ mm and 150 mm.
The 150 mm monitor is only used for the 174\degree mode optical path split before the IP.
The 100 mm monitor accepts the 174\degree mode path guided after the IP,
and paths from all lower angle modes.
}
\label{fig:monitor-location}
\end{center}
\end{figure}

For the Shintake monitor, we need to stabilize the fringe phase at the IP.
Since we cannot install the phase monitor at the IP,
we have to install the monitor at other locations.
When we control the fringe phase viewed at the monitor location, 
the phase difference between the IP and the monitor location
causes a phase error at the IP.

To minimize the error, two phase monitors are installed in 174\degree crossing angle mode.
As shown in Fig.~\ref{fig:monitor-location}, one of the monitors,
which is located at 150 mm height from the table surface, views the relative phase
of the laser beams separated from the main laser beam lines before the IP (green line),
and the other monitor, located at 100 mm height, views the laser beams
which have passed through the IP (red line).
Since the IP is between the two monitoring position, the phase at the IP is
considered to be between measured phases by the two monitor.
The two monitors are used to improve the phase accuracy at the IP
and to estimate the effectiveness of the phase stabilization at the IP.

For other angle modes, only one phase monitor is active because of geometrical restrictions.
Since the accuracy is less important in these angle modes,
it is not seriously concerned if the phase accuracy is not assured.

\subsection{Tests of Phase Monitoring and Stabilization}
\label{sec:teststab}

\subsubsection{Measurement Condition}

To obtain a performance of the phase monitoring and the stabilization, we did test experiments
of the phase monitoring and the control. For the phase control test, a couple of objective 
lenses with image sensors, and a delay line with a piezo stage, as all described in former sections, are used.
We split the laser beams and form two beam crossings to be viewed by the phase monitors (Ch1 and Ch2),
as same as 174\degree setup of the real Shintake monitor.
The two monitors are used to estimate the performance of the phase stabilization.

To estimate the performance of the phase stabilization for the Shintake monitor,
the following method is applied.
\begin{itemize}
	\item Obtain and record phase data from both monitors.
	\item Feedback routine calculates the motion of the phase mover
		to stabilize the Ch1 phase. Ch2 phase is recorded but not used for the stabilization.
	\item Analyze the phase stability of Ch2 phase.
		The Ch2 phase stability indicates the stability at the IP of the Shintake monitor for the real setup.
\end{itemize}

\begin{figure}
\begin{center}
\includegraphics[width=25em]{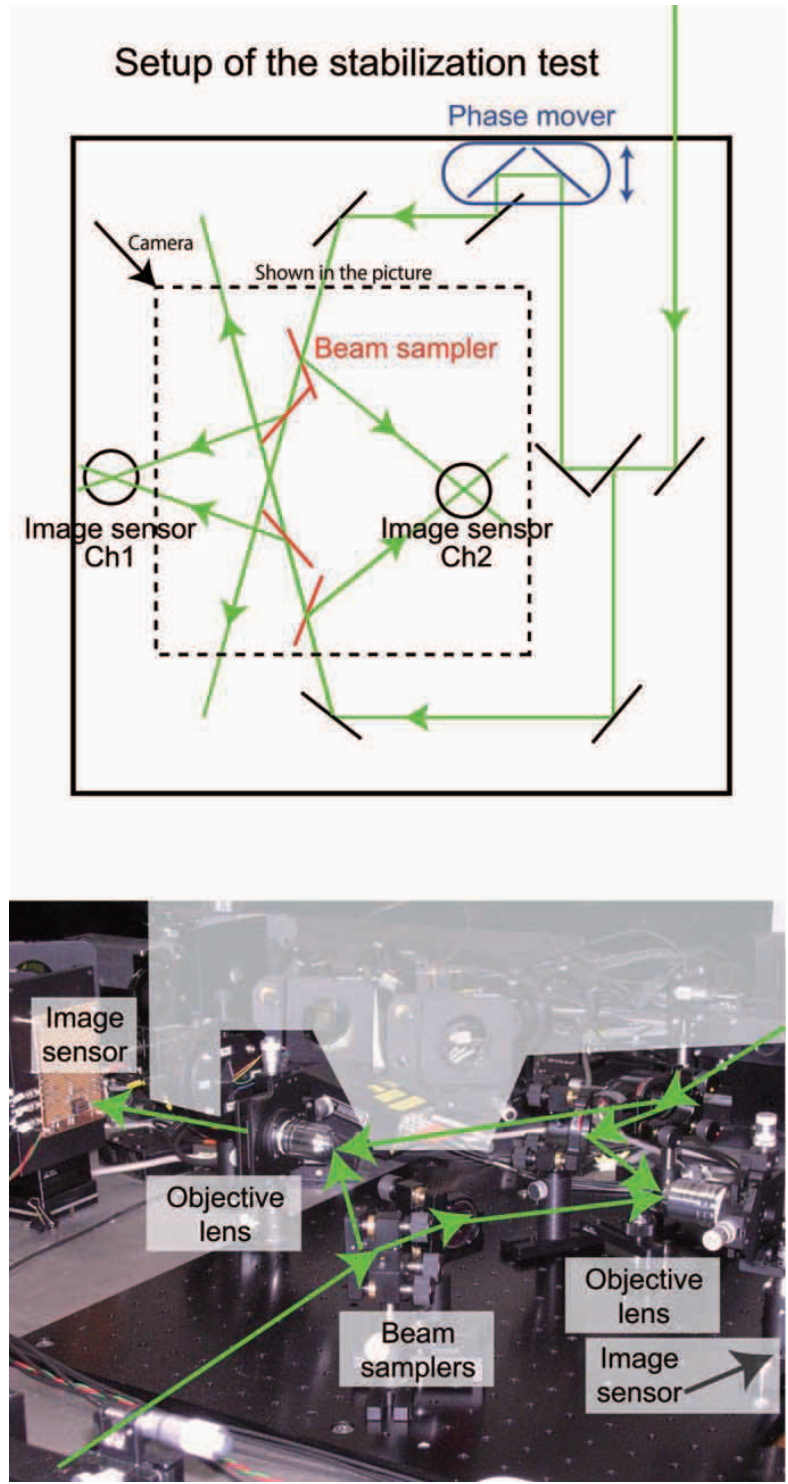}
\caption{Upper: the layout of the stabilization test.
Lower: a picture of the test setup of the phase monitoring.}
\label{fig:test-picture}
\end{center}
\end{figure}

Figure \ref{fig:test-picture} shows the setup of the experiment.
The experiment is performed on the existing optical table
used in the FFTB Shintake monitor with additional instruments.
The stabilization measurement is performed using the pulsed laser to be used in
the real ATF2 Shintake monitor,
and a low power continuous wave (CW) laser for a comparison.

\subsubsection{Result of the CW Laser Test}

\begin{figure}
\begin{center}
\includegraphics[width=30em]{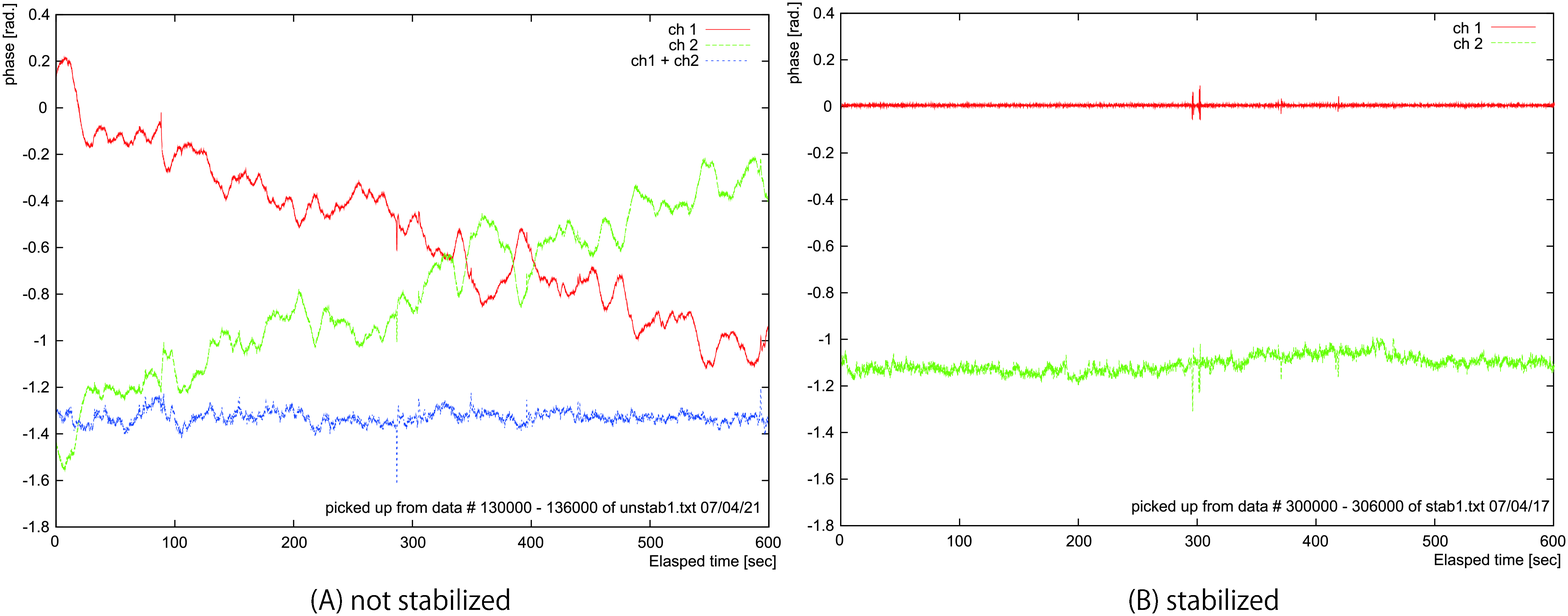}
\caption{A sample result of the phase monitoring in a 10 minutes window
(A) without phase stabilization and (B) with phase stabilization.
In (A), an addition of both channels is shown in the blue line.}
\label{fig:phase-cw}
\end{center}
\end{figure}

Figure \ref{fig:phase-cw} (A) shows a typical result of the phase
monitoring without the phase stabilization.
Because the phase monitors are located face-to-face, 
phase variation can be observed inversely to the other.
The anti-correlation of two monitors can be seen by adding two data (the blue line).
It shows a strong anti-correlation on the 10 minutes window in the figure.
The anti-correlation factor is expected to be the relative phase fluctuation
between the incident two laser beams.
The data of individual channels show a fluctuation of relatively short period
of $\lesssim$ 1 minute and a long term drift over the 10 minutes window.

Figure \ref{fig:phase-cw} (B) shows a typical result of the
phase monitoring with the phase stabilization.
Ch1 is stabilized to phase = 0, which results in a very flat plot of the red line.
Ch2 is not stabilized, but because the stabilization is performed before
splitting the laser path, Ch2 is also affected by the stabilization.
The phase variation of Ch2 is drastically suppressed compared to the
unstabilized data, while the pulse-to-pulse fluctuation seems to increase slightly
due to the phase stabilization.

For our modulation measurement,
stability in 1 minute (time for single measurement) is concerned.
To obtain 1 minute stability, we sliced the phase data
to 1 minute windows and acquire RMS values in the window.

The averages of the Ch2 RMS phase fluctuations in 1 minute windows are
27.5 mrad.~with stabilization of Ch1 (corresponding to the green line in Fig.~\ref{fig:phase-cw} (B)),
and 73.6 mrad.~without stabilization (the green line in Fig.~\ref{fig:phase-cw} (A)).
A clear effect of phase stabilization is observed.

\subsubsection{Result of the Pulsed Laser Test}

\begin{figure}
\begin{center}
\includegraphics[width=30em]{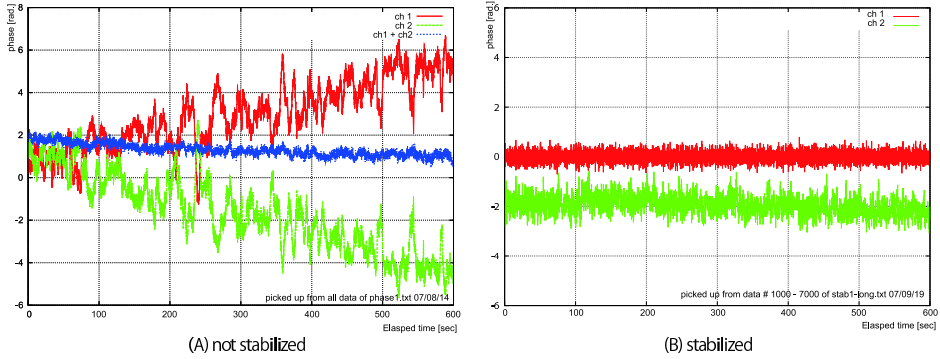}
\caption{A sample result of the phase monitoring in a 10 minutes window
(A) without phase stabilization and (B) with phase stabilization in the pulsed laser.
The vertical axis is enhanced compared to Fig.~\ref{fig:phase-cw}.}
\label{fig:phase-pulse}
\end{center}
\end{figure}

For the pulsed laser, we perform the same measurement as the CW laser.

Figure \ref{fig:phase-pulse} (A) shows a typical result of the phase
monitoring for the pulsed laser.
A pulse-to-pulse fluctuation of around 0.5 radian amplitude can be clearly observed,
in addition to the fluctuation of multi-second timescale which can be seen also in the CW laser.
The pulse-to-pulse fluctuation can also be observed in the spectrum of adding two channels (blue).
Because the fluctuation is pulse-to-pulse and correlation between channels is poor,
this fluctuation cannot be canceled nor corrected by the phase monitoring and stabilization
system.
The RMS phase of the addition spectrum is 170 mrad., which is the limit of the phase stabilization
performance for the pulsed laser.
For the long term fluctuation two channels are almost correlated and canceled by adding the two
channels.

Figure \ref{fig:phase-pulse} (B) shows a 10 minutes spectrum of 
a stabilized phase data of the pulsed laser. The stabilization is effective for long term
fluctuations, but the pulse-to-pulse fluctuation is totally remained or slightly enhanced.

The obtained RMS stabilities of 1 minute windows for the pulsed laser are
239 mrad.~with stabilization of Ch1 and 800 mrad.~without stabilization.
Although the stability is quite worse than the stability in the CW laser,
it is still in an acceptable range for the target performance of the Shintake monitor.
The stability is limited by the pulse-to-pulse fluctuation.
The pulse-to-pulse fluctuation seems to be partly caused by the angular jitter,
while further study is needed to understand and suppress the pulse-to-pulse fluctuation.

\subsection{Beam Position Stability and an IP-BPM}

Electron beam position jitter with respect to the optical table is also a source
of the phase jitter.
To monitor the electron beam position jitter, a ultra-high precision cavity beam position monitor
(IP-BPM) is attached to the optical table of the Shintake monitor.
The vertical position resolution of the IP-BPM is demonstrated in ATF and
8.7 nm resolution is already obtained.
Performance study of the IP-BPM is described elsewhere\cite{ipbpm}.

Design beam position stability at the IP in ATF2 is $1/3\sigma_y$, about 12 nm for
37 nm beam size.
Although the actual position stability is unknown,
we can correct the electron beam position using the IP-BPM with 8.7 nm resolution.

\subsection{Summary}

In summary, current obtained phase accuracy is about 240 mrad.~(corresponding to 10.1 nm)
by phase stabilization, and 8.7 nm by IP-BPM.
Combining these, 13.3 nm (0.31 radian) is the current estimated phase stability
obtained in the real Shintake monitor.
For the performance estimation, this 13.3 nm is used for the phase uncertainty in the beam size measurement.
This value may be lowered by suppressing pulse-to-pulse fluctuation at the phase monitor,
by improving air-flow prevention, and/or suppressing noise of the IP-BPM.


\section{Fringe Contrast}
\label{sec:contrast}

Contrast of the interference fringe strongly affects modulation measurements.
We need to ensure the fringe contrast is perfect, or to know the contrast precisely
if it is not perfect.
In this section we briefly overview contrast degradation sources,
and discuss about the contrast estimation.

\subsection{Sources of Contrast Degradation}
\label{sec:contrast:theo}

\begin{itemize}
\item Power Imbalance

Power imbalance of the crossing two laser beams causes the contrast degradation.

The modification of the modulation depth $M$ by the power imbalance $P_I = P_1/P_2$ is given by
\begin{equation}
	M' = \frac{2\sqrt{P_I}}{1+P_I}M.
	\label{eqn:contrast:moddepth-power2}
\end{equation}

In our design, the power imbalance is estimated to be $<$ 12\%.
Since the calculated modulation degradation by (\ref{eqn:contrast:moddepth-power2})
of the 12\% power imbalance ($P_I = 0.88$) is only 0.2\%,
the effect of power imbalance can be ignored.

\item Position Displacement

Position displacement causes a localized power imbalance and then
degrade the fringe contrast.
If our design of the position adjustment and the image inversion feature
(discussed in Section \ref{sec:laserpos}) are correctly worked,
the position displacement between two laser beams should be $<$ 1 $\mu$m.
Using (\ref{eqn:contrast:moddepth-power2}),
the contrast degradation is estimated to be $<$ 0.3\%.

\item Imperfect Polarization

Since photons of different polarizations are not interfered,
imperfect polarization of the laser photons cause the modulation degradation.

We can estimate the modulation degradation by the limited polarization ratio $P (<1)$ as
\begin{equation}
	M' = \{P^2+(1-P)^2\}M.
\end{equation}

The polarization factor of our laser is measured to be $>$ 99.3\%.
It corresponds to $<$ 0.5\% modulation degradation.

\item Spherical Wavefront

The focusing laser beams have spherical wavefronts.
Wavefront radius $R$ is determined by Gaussian beam optics as
\begin{equation}
	R = z\left\{1+\left(\frac{\pi{}w_0^2}{\lambda{}z}\right)^2\right\}
\end{equation}
where $z$ is distance along the beam line from the waist point,
$w_0$ is the waist beam size and $\lambda$ is the laser wavelength. 
If the electron beam crosses just on the focal
point of the laser beam ($z=0$), the $R$ comes infinite and the wavefront is perfectly
planar, which can make planar interference fringes.
Practically, the electron beam goes through somewhat distant position from the
focal point, and the $R$ comes finite value, which stands for the imperfect planar
interference fringes and consequently it results modulation degradation.

The modulation degradation is given by
\begin{equation}
	\frac{M'}{M} = \int^{+\infty}_{-\infty}\exp\left(\frac{\rho^2}{w_z^2}\right)
									\sin\left(\frac{k\rho^2}{2R}-\phi\right)d\rho
\end{equation}
where $\rho$ is the position perpendicular to the beam axis ($\rho = 0$ as beam center),
$w_z$ is the beam size at $z$, and
$\phi$ is the phase determined to make the integral maximum.

We assume that the accuracy of the focal length alignment
can be achieved to be $<$ 400 $\mu$m, which causes
0.6\% contrast degradation.

\item Spatial Coherence

If the spatial coherence of the laser beams is poor,
the formed interference fringes are distorted
at the laser beam tail, and the contrast is degraded.

The spacial coherence become poor if the mirrors or lenses are distorted,
especially those in laser paths after the main beamsplitter.
We use optical components which have $1/10\lambda$ or better quality to suppress
the coherence degradation.
In addition, overlapping of higher-order paths, eg.~by back reflection of mirrors,
may degrade the spatial coherence, but the effect is suppressed around the focal point
because the higher-order paths usually have different focal points.

Though the spatial coherence is difficult to be directly measured,
we expect that it is good enough not to degrade the fringe contrast.

\item Temporal Coherence

Temporal coherence (also known as coherent length) is one of the basic characteristics
of the laser. The laser light outside the coherent length is neither coherent nor interfered.
The temporal coherence is determined by the wavelength width of the laser oscillation.
Using the laser specifications of the spectral width $\delta{}k < 0.003$ cm$^{-1}$
and the difference of the split path lengths of the Shintake monitor $\Delta{}L < 100$ mm,
the maximum phase variation is 
\begin{equation}
	\delta{}\phi = \delta{}k\Delta{}L = 3 \times 10^{-2} \ \mathrm{[rad.]}.
\end{equation}
which results in $<$ 0.6\% contrast degradation.

\end{itemize}

All above contrast degradation effects but the spatial coherence
are shown to affect the contrast less than 1\% level.
We expect that the overall contrast degradation at the IP is less than 10\%.
Practically we need to measure the fringe contrast in the real setup
and correct the beam size calculation using the contrast value.

\subsection{Contrast Measurement}

\begin{figure}
\begin{center}
\includegraphics[width=15em]{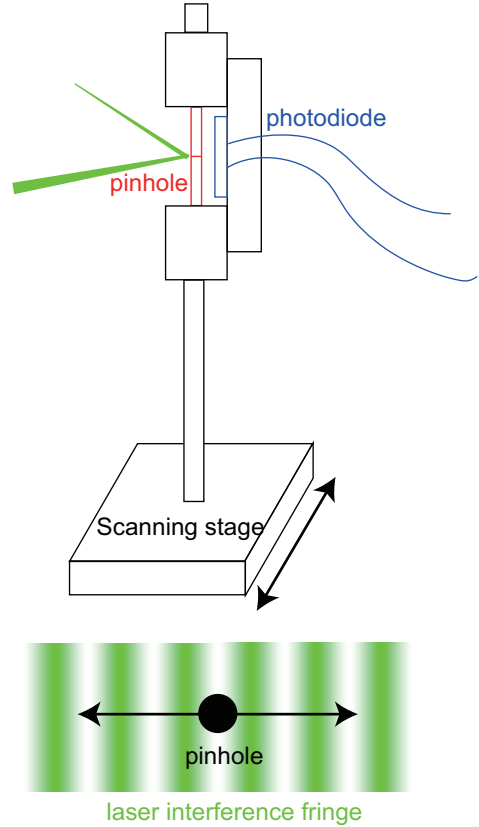}
\caption{
A schematic of a pinhole fringe monitor.
Laser light passing through the pinhole is captured by a photodiode.
A motorized stage is equipped to sweep the pinhole across interference fringes.
}
\label{fig:pinhole-schematic}
\end{center}
\end{figure}

Contrast of the interference fringe is measured using pinholes in a test setup.
Figure~\ref{fig:pinhole-schematic} shows a schematic drawing of
the pinhole monitor. In the monitor, a pinhole is placed close to a photodiode,
attached to a stage with a motorized actuator. Split laser beams cross at the
surface of the pinhole to make interference fringe pattern.

We measure interference fringes formed by both CW and pulsed lasers
using this pinhole monitor.
For the CW laser, 1 $\mu$m aperture pinhole was used with the crossing angle
of 6\degree.
For the pulsed laser, 1 $\mu$m pinhole cannot be used because of heat destruction,
thus high-power durable 5 $\mu$m pinhole was used instead with 2\degree crossing angle.
Additional contrast degradation in the measurement due to the finite pinhole diameter
was compensated by calculations using the pinhole diameters.
Relatively large uncertainty of the pinhole diameter (in specification) was treated
as a systematic error.

\begin{figure}
\begin{center}
\includegraphics[width=30em]{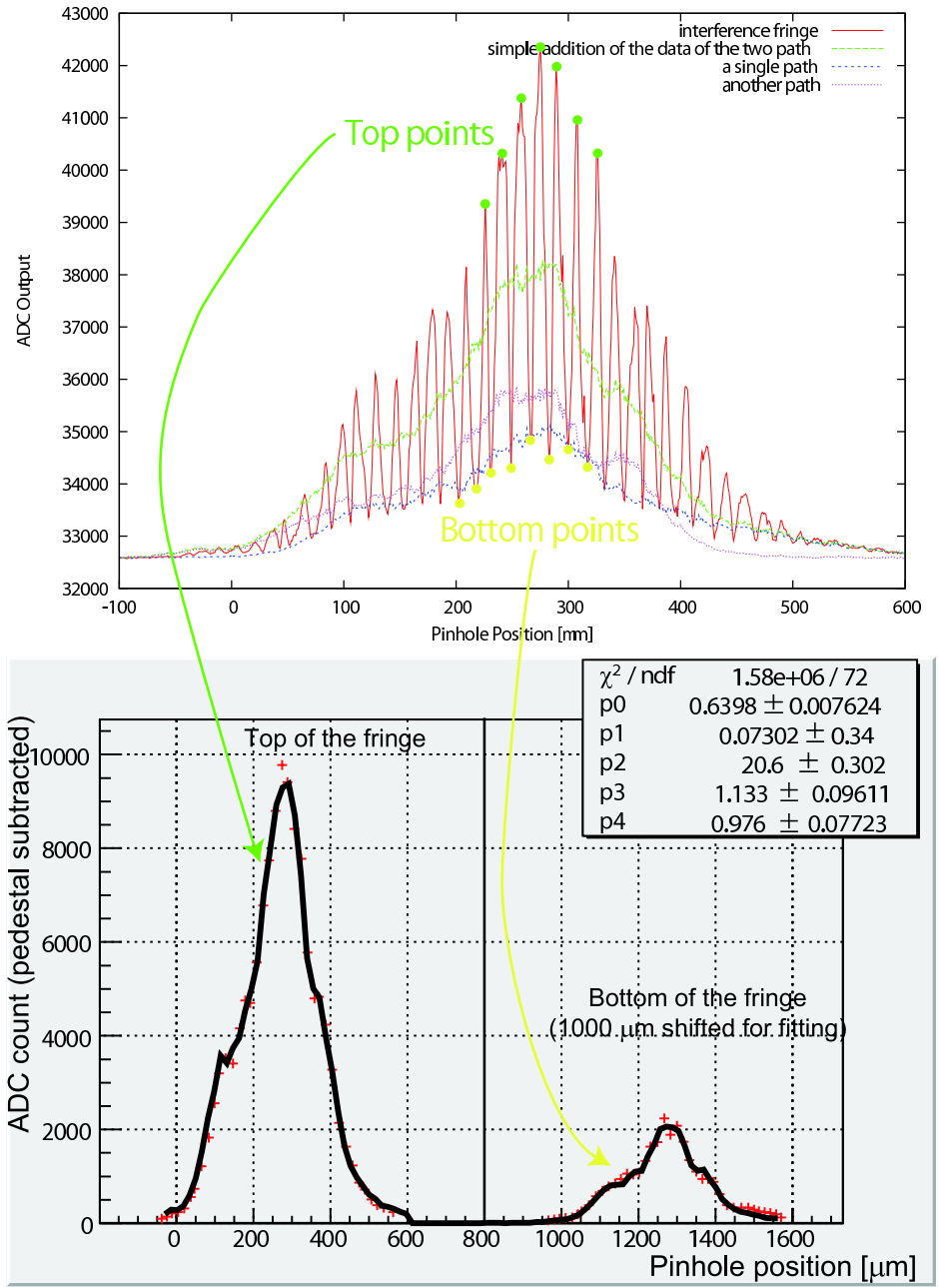}
\caption{A contrast measurement of the pulsed laser. Upper: Observed fringe pattern
by intersecting two laser paths. The red and blue lines show the intensity of individual paths.
Lower: A result of contrast calculation. The top and bottom points are picked up from
the fringe pattern, and simultaneously fitted by shifting the bottom points 1000 $\mu$m.
The fitting function is described in the text.
Free parameters of the fitting are:
fringe contrast (p0), position shift of each path (p1 and p2), and power shift of 
each path (p3 and p4).}
\label{fig:pulse-contrast}
\end{center}
\end{figure}

Figure \ref{fig:pulse-contrast} shows the result of the measurement with the pulsed laser.
To obtain the fringe contrast, following method is used.

\begin{enumerate}
	\item The fringe pattern is formed by intersecting two laser paths.
	We define the obtained fringe pattern as $f_\mathrm{orig}(x)$.
	The pinhole sweeping is also performed for each laser path open and another closed.
	The obtained distributions are denoted as $f_1(x)$ and $f_2(x)$.
	\item The peak and bottom points are extracted from $f_\mathrm{orig}(x)$.
	The group of the top points are denoted as $f_\mathrm{max}(x)$,
	and the group of the bottom points are denoted as $f_\mathrm{min}(x)$.
	\item A distribution for the fitting $f(x)$ is defined as,
	\begin{equation}
		f(x) \equiv \left\{
			\begin{array}{ll}
				f_\mathrm{max}(x) & (x < x_t) \\
				f_\mathrm{min}(x - x_0) & (x \geq x_t)
			\end{array}\right.
	\end{equation}
	In Fig.~\ref{fig:pulse-contrast}, $x_0 = 1000$ [$\mu$m] and $x_t = 800$ [$\mu$m].
	\item Using fitting parameters $p_0$ - $p_4$, the fitting function $g(x)$ is defined by,
	\begin{equation}
		g(x) = \left\{
			\begin{array}{l}
				p_0\left(\sqrt{p_3f_1(x - p_1)} + \sqrt{p_4f_2(x - p_2)}\right)^2 \\
					\qquad\qquad + (1 - p_0)(p_3f_1(x - p_1)) + p_4f_2(x - p_2)) \quad (x < x_t) \\
				p_0\left(\sqrt{p_3f_1(x - p_1 - x_0)} - \sqrt{p_4f_2(x - p_2 - x_0)}\right)^2 \\
					\qquad\qquad + (1 - p_0)(p_3f_1(x - p_1 - x_0) + p_4f_2(x - p_2 - x_0)) \quad (x \geq x_t)
			\end{array}\right.
	\end{equation}
	The points of the functions where no data points are provided are defined by interpolations or
	extrapolations of the nearest two points.
	$p_0$ means the fringe contrast (0-1), $p_1$ and $p_2$ means the horizontal shifts of the
	$f_1(x)$ and $f_2(x)$, $p_3$ and $p_4$ means the power fluctuation of the
	$f_1(x)$ and $f_2(x)$.
	The initial values of $p_1$ and $p_2$ are manually given since the optimization of them is
	not efficient by the fitting.
	For the CW laser measurement, $p_1$ is set to 0.005 [mm] and $p_2$ is set to 0.
	The movable range of $p_3$ and $p_4$ is limited to $1 \pm 0.05$.
	\item Using $p_0$ as ``the fitted fringe contrast".
	From Fig.~\ref{fig:pulse-contrast}, $64.0 \pm 0.8$\% contrast is obtained by the fitting.
\end{enumerate}

\begin{table}
	\begin{center}
		\begin{tabular}{|c|r|r|r|r|}\hline
			Laser & Angle & Pinhole Aperture & Contrast (raw) & Contrast (compensated) \\ \hline
			CW & 6\degree & $1.0 {+0.4 \atop -0.0}$ $\mu$m & $70 \pm 2.0$\% & $75 \pm 2.0 \mathrm{(fit)} \pm 2.5 \mathrm{(aperture)}$ \\ \hline
			Pulsed & 2\degree & $5.0 \pm 0.75$ $\mu$m & $64 \pm 0.8$\% & $75 \pm 0.8 \mathrm{(fit)} \pm 4.0 \mathrm{(aperture)}$ \\ \hline
		\end{tabular}
	\caption{Summary of the contrast measurements by pinholes.}
	\label{tbl:pinhole}
	\end{center}
\end{table}

Table \ref{tbl:pinhole} shows a summary of the contrast measurements.
Both lasers give consistent results,
and they show that contrast around 75\% can be acchievable.
Compared to the estimation described in Section \ref{sec:contrast:theo},
however, unexpected contrast degradation is measured and the reason is
currently unknown.
Since contrast in the real setup may be differ from the test setup.
We can more reliable contrast measurements using electron beam in the real setup

\subsection{Contrast Estimation by Beam Size Measurements}
\label{sec:contrast:beam}

Since the offline contrast measurement does not
have enough accuracy, contrast estimation using the electron beam is necessary.
The following methods are only applicable after the ATF2 beam is available.

\subsubsection{Contrast Estimation of Small Crossing Angles}

The contrast estimations of the small crossing angle modes are relatively easy
if the small electron beam size is obtained.
When the smaller beam size than the measurable range is arrived,
the obtained modulation depth is $\sim$ 100\%.
Then, if the fringe contrast is degraded, the degradation is directly observable
by the measurement.
Practically, minimum fringe contrast is obtained by measuring the same beam size
by two crossing angles, using a relation shown in Fig.~\ref{fig:moddepth}(right).

In this method, the estimation of the modulation degradation is efficient 
if the electron beam size is reduced to the efficient range of 
the larger crossing angle modes.
In different words, the estimation of the 2\degree mode needs 
$\lesssim 1$ $\mu$m electron beam size, that of the 8\degree mode needs
$\lesssim 300$ nm beam size, and that of the 30\degree mode needs
$\lesssim 70$ nm beam size.
Degradation of the 174\degree mode cannot be estimated by this method
because there are no larger crossing angle modes and
the modulation depth cannot be high enough because the ATF2 cannot supply
such a small beam size.

However, if the obtained modulation degradation is almost zero for the rest of
the modes, or at least the degradation is almost the same for the modes,
the degradation of the 174\degree mode is assumed to be mostly the same.
Since this method can accurately acquire the degradation for the smaller angle modes,
it can be a help to assume the degradation of the 174\degree mode.

\subsubsection{Estimation of the 174\degree Mode}

The previous method cannot estimate the modulation degradation
of the 174\degree mode (but only can `assume' the degradation).
If we can acquire several beam sizes whose ratio can be estimated without data
of the Shintake monitor, we can acquire the modulation degradation
by measuring the beam sizes.

The ratio of the modulation depth of two beam sizes can be written as,
\begin{eqnarray}
	\frac{M_1}{M_2} &=& \frac{M_\mathrm{deg}|\cos2\phi|\exp[-2(k_y\sigma_{y1})^2]}
													{M_\mathrm{deg}|\cos2\phi|\exp[-2(k_y\sigma_{y2})^2]} \\
	&=& \exp[-2k_y^2(\sigma_{y1}^2 - \sigma_{y2}^2)] \\
&=&	\exp\left[-2k_y^2\sigma_{y1}^2\left\{\left(\frac{\sigma_{y2}}{\sigma_{y1}}\right)^2-1\right\}\right]
\label{eqn:contrast:bsratio}
\end{eqnarray}
where $M_\mathrm{deg}$ is the contrast degradation factor.

If we know the ratio of the two beam sizes $\sigma_{y2}/\sigma_{y1}$ and using 
the measured ratio of the modulation depth $M_1/M_2$,
we can obtain $\sigma_{y1}$ by (\ref{eqn:contrast:bsratio}).
Since this method is independent of the modulation degradation $M_\mathrm{deg}$,
we can acquire the $M_\mathrm{deg}$ by the normal calculation (\ref{eqn:moddepth})
using $\sigma_{y1}$ obtained by this method. The $M_\mathrm{deg}$ is thought not to be
fluctuated time by time, so if we measure $M_\mathrm{deg}$ once,
the obtained value should be used for a long period.

To perform this measurement, the beam size ratio $\sigma_{y2}/\sigma_{y1}$ must be needed.
One possibility is to change the electron beam emittance.
Since the electron beam size is proportional to $\sqrt{\mathrm{emittance}}$ everywhere
in the beam line, the emittance ratio between two electron beams are easily measured by
measuring beam sizes by traditional BSMs where the beam size is much larger than the IP.
The beam size larger than 10 $\mu$m can be measured with accuracy around 1\%.

If we assume that the 50 nm and 75 nm beam size with relative beam size known by
1.4\% accuracy (corresponding to 1\% for each), the estimation error of the $M_\mathrm{deg}$
comes 3.5\%, which is in allowable range of the Shintake monitor.


\section{Background in the Gamma Detector}
\label{chap:detector}

Background of the gamma detector is crucial for the modulation
resolution. Estimation of assumed background, its effect on the
measurement, and strategy to suppress the background
are discussed in this section.

\subsection{Background Sources}

\subsubsection{Synchrotron Radiation from Final Focusing Magnets}

Electron beam passing through final focusing magnets
is the main source of the synchrotron radiation for our detector.
Energy spectrum of synchrotron radiation is characterized by ``critical energy",
which is 1.1 keV in ATF2 assuming 1 Tesla magnetic field.
Since energy distribution of the synchrotron radiation photons is 
limited up to several times the critical energy,
all of the photons from synchrotron radiation must be stopped at the beam pipe
(1.6 mm thick made of SUS304) and do not arrive at the detector.

\subsubsection{Beam Scattering with Residual Gas}

Since the vacuum in the accelerator beam pipe is not perfect, some of the electron
beam can be scattered by residual gas in the beam pipe. The amount of scattering
must be estimated. Since Rutherford scattering is adiabatic and does not emit
photons, we need to consider only bremsstrahlung process. 

Cross section of the bremsstrahlung of an electron interacting with the residual gas
for $> 1.3$ keV photon emission (fraction of $1 \times 10^{-6}$ from the beam energy)
 is estimated to be
\begin{equation}
	\sigma_b = 2.5 \times 10^{-22} [\mathrm{cm}^2]
\end{equation}
assuming N$_2$ gas\cite{handbook-gasscatter}.

The critical section for the photon emission is between the last bending magnet
before the final focus area and the bending magnet after the IP ($L \sim 1200$ [cm]),
and pressure in the ATF2 beam line is lower than $10^{-6}$ [Pa], corresponding to
gas density of $n_\mathrm{gas} = 2.7 \times 10^8$ [particles/cm$^3$].

The calculated number of scattered photons per bunch ($n_e = 1 \times 10^{10}$) is,
\begin{equation}
	n_\gamma = \sigma_bLn_\mathrm{gas}n_e \sim 0.8
\end{equation}
which is negligible because number of signal photons is over $10^3$ per bunch.

\subsubsection{Beam Halo Scattering with Beam Pipe}

Charge distributions of accelerator beams can be separated to two parts,
beam cores which usually have Gaussian-like distributions,
and beam halos which have much broader distributions than beam cores.
Diameters of beam transportation pipes are designed so that 
the beam core does not hit the beam pipes.
However, in some parts of ATF2 beam line, beam size is enhanced by the optics and a part of
the beam halo can hit the beam pipes, which emit bremsstrahlung photons.

Energy distribution of this background is already shown in Fig.~\ref{fig:gammaenergy}.
Since photon energy of this background is very large,
we need to consider effects of the background seriously
if amount of the background is not negliglble.
To estimate this background, we measured charge distribution of a beam halo
in current ATF extraction line. The measurement result is described in the
next subsection.

\subsubsection{Particles from Beam Dump}

Since a beam dump is located close to the IP in the ATF2, it can be a large
background source. Amount of detected background from the beam dump depends largely on 
the detector location. 
Since most of particles from the beam dump are backscattered ones to the 
upstream of the beam line, we can install the detector behind the beam dump
to suppress this background. 

\subsection{Distribution of the Beam Halo}

Understanding charge distribution of the beam halo is essential to our estimation
of background. As the construction of the ATF2 beam line is not completed yet,
we measured the distribution at current ATF extraction line which is located
upstream of the new ATF2 line. The measurement was performed in ATF spring run
of 2005.

\subsubsection{Measurement Setup}

\begin{figure}
\begin{center}
\includegraphics[width=30em]{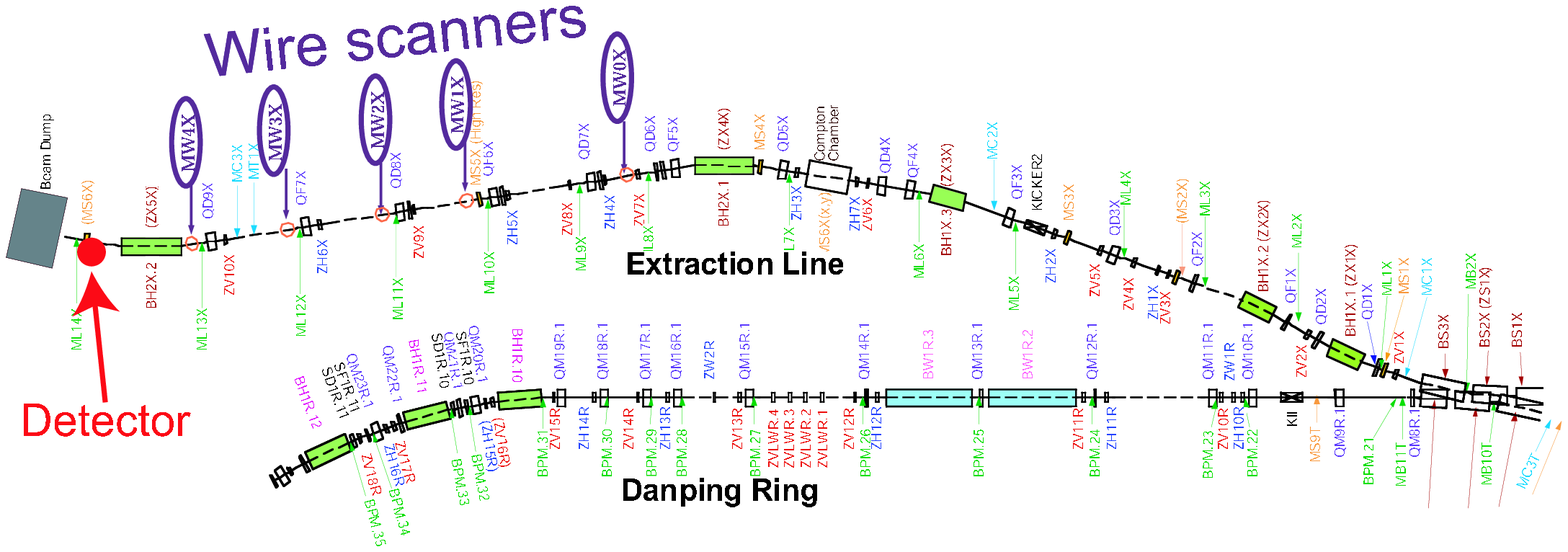}
\caption{Location of wirescanners and their detector in ATF extraction line.}
\label{fig:detector:halomeas-location}
\end{center}
\end{figure}

We use wirescanners installed in the ATF2 extraction line for the scattering target.
The wirescanner consists of a metal wire (tungsten in the ATF)
with a micromover to scatter the electron beam at every wire position.
The scattered photons are counted by a gamma detector,
which is an air-Cherenkov counter with a 2 mm thick lead converter
and a PMT is attached for the photon counting.

Figure \ref{fig:detector:halomeas-location} shows the location of wirescanners and
a gamma detector. Five wirescanners are installed, while all of them are read
by a single detector.
Since electron beam size at each wirescanner is varied, we use
all detectors to obtain the halo charge distribution of the various beam size.

\subsubsection{Measurement Result}

\begin{figure}
\begin{center}
\includegraphics[width=25em]{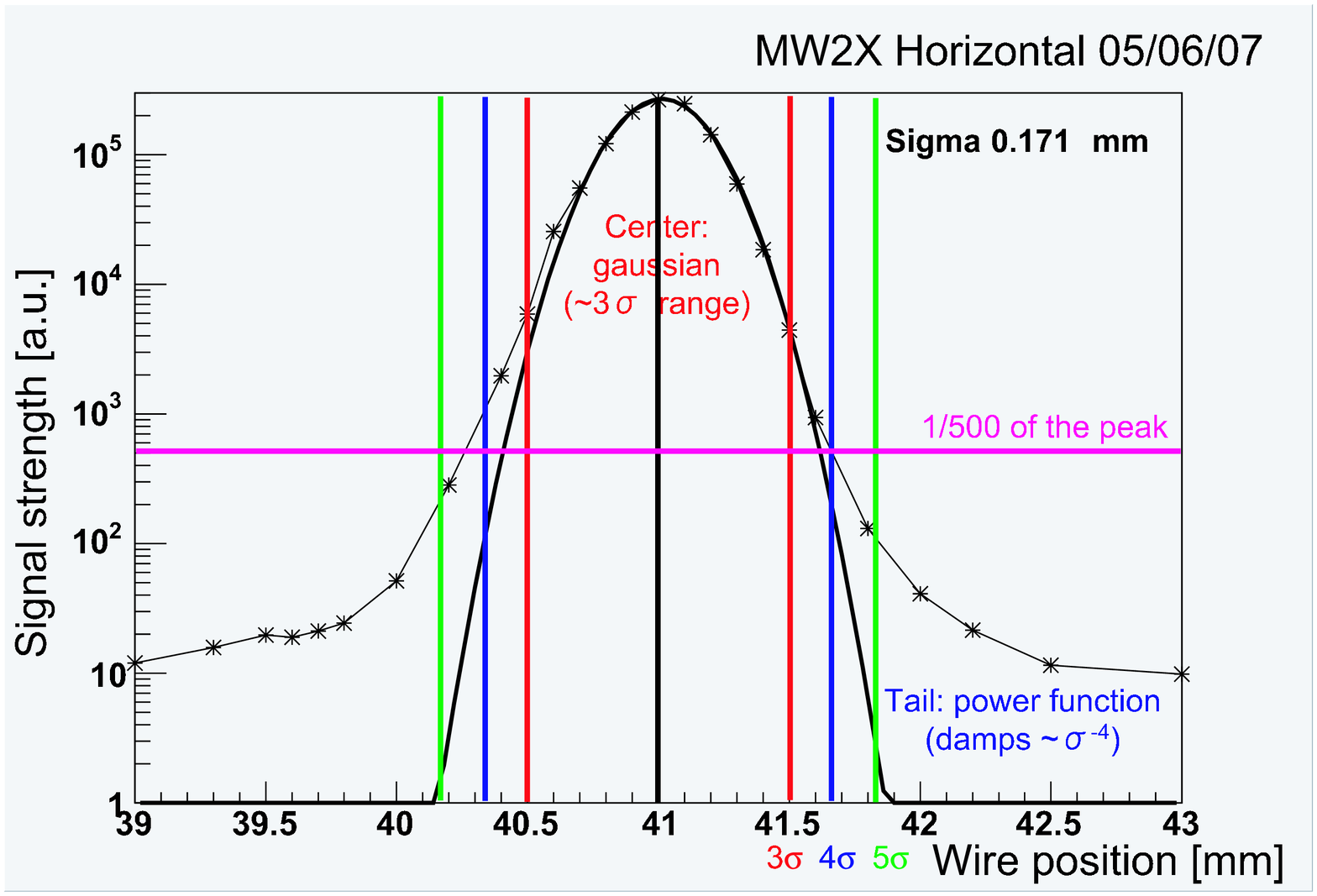}
\caption{Measured result of charge distribution using an ATF extraction line wire scanner.}
\label{fig:detector:halo-center}
\end{center}
\end{figure}

\begin{figure}
\begin{center}
\includegraphics[width=30em]{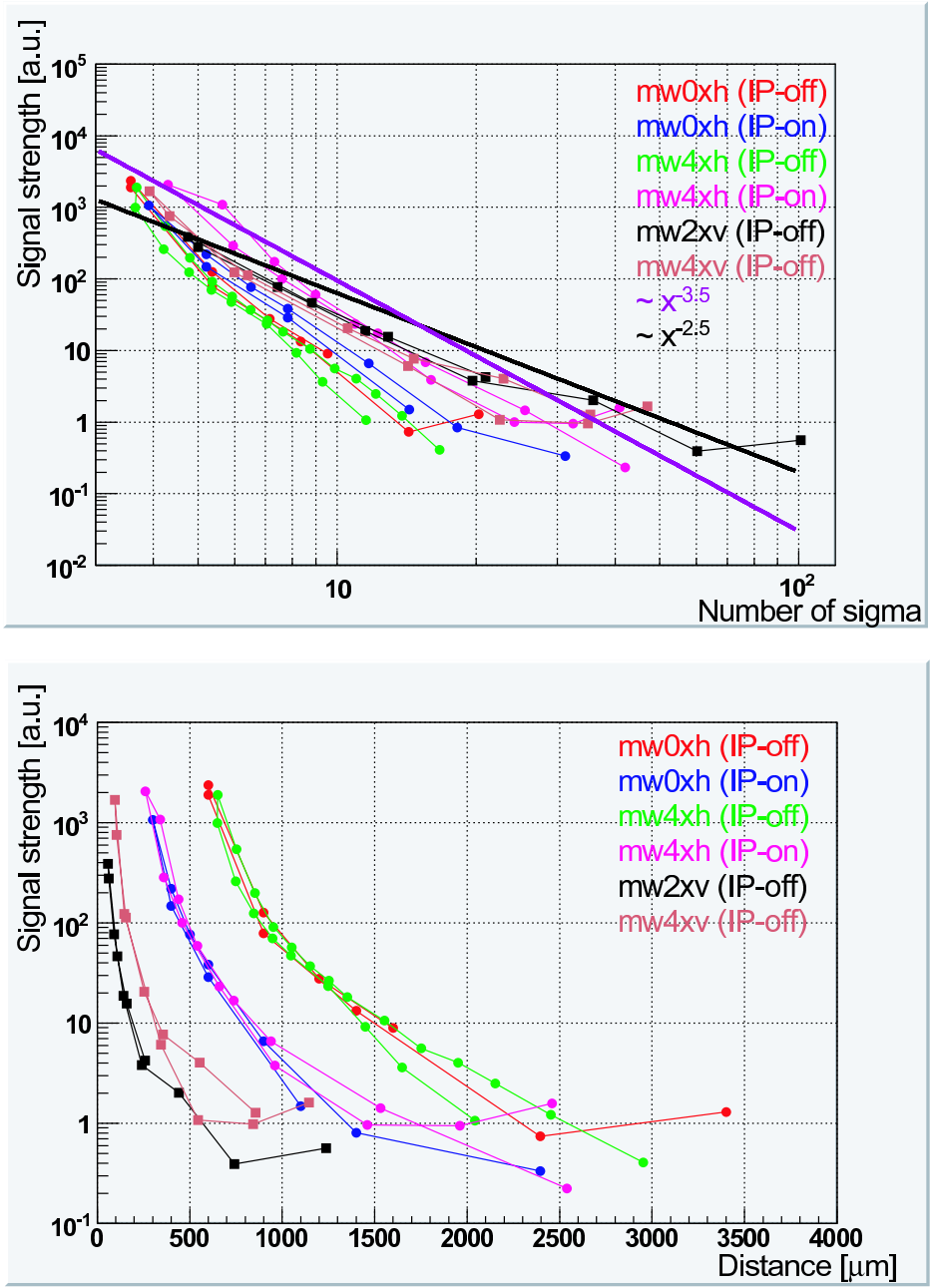}
\caption{Measurement of the halo part using several wirescanners for both vertical and horizontal
directions. Upper graph shows the plot in which horizontal axis is normalized by its beam size.
Lower graph shows the plot in which horizontal axis is just a distance from the beam center.
For both graphs, vertical beam profiles are shown as a square, and horizontal beam profiles
are shown as a circle. The difference of the IP-on data and the IP-off data is the vacuum
level. For the IP-off data, some of the ion pumps in the ATF dumping ring were turned off
to obtain data with degraded vacuum. The difference of the vacuum level is about 1:5.}
\label{fig:detector:halo-sigma-distance}
\end{center}
\end{figure}

A result of the measurement is shown in Fig.~\ref{fig:detector:halo-center} and
Fig.~\ref{fig:detector:halo-sigma-distance}.
Figure \ref{fig:detector:halo-center} shows a measurement of charge distribution of
wide dynamic range. An applied voltage of PMT is varied to obtain wide dynamic range.
The plot shows that
the distribution in the beam center of $< 3\sigma$ range is well approximated by a Gaussian
(bold line), while in the region of $> 3\sigma$, the deviation from the central Gaussian is large.
Figure \ref{fig:detector:halo-sigma-distance} shows a comparison of the halo distribution
for several beam sizes. In the upper graph horizontal axis is in a unit of beam size, and 
in lower graph it is in a absolute distance. Comparing two graphs, we can understand
a width of the halo expansion is almost proportional to the center beam size.

In the upper graph of the Fig.~\ref{fig:detector:halo-sigma-distance},
two straight line are also shown. The blue line shows a function
of the $-$3.5th power of the distance from the center,
and the black line shows a function of the $-$2.5th power.
According to the graph, for the horizontal distribution $-$3.5th power
is more likely to the data, while for the vertical distribution $-$2.5th power
is more likely for $>$ 6 $\sigma$ region.

\subsubsection{Estimation of Amount of Beam Halo}

We can estimate the maximum charge distribution of the beam halo using the halo measurement.
We assume that the halo distribution of both vertical and horizontal
follow to the $-$3.5th power function until 6 $\sigma$,
and for outside 6 $\sigma$ the vertical distribution follows to the $-$2.5th power function
while the horizontal distribution still follows to the $-$3.5th.
The ratio of the charge intensity between the central Gaussian and halo function is determined so
that the density of the center is 500 times larger than the density at the 5 $\sigma$ position.
This value is extracted from Fig.~\ref{fig:detector:halo-center}.

Since the most of the charge in a bunch is concentrated in the central Gaussian, 
density at the center $\rho_c$ can be easily obtained as
\begin{equation}
	\rho_c = \frac{N}{\sqrt{2\pi}\sigma},
\end{equation}
where $\sigma$ is the beam size of the central Gaussian and $N$ is the bunch population, which is
assumed to be $1 \times 10^{10}$ electrons in this calculation.
Since we assume the density of the 5 $\sigma$ is 1/500, which corresponds to $8.0 \times 10^6$ divided
by the $\sigma$.

\begin{figure}
\begin{center}
\includegraphics[width=25em]{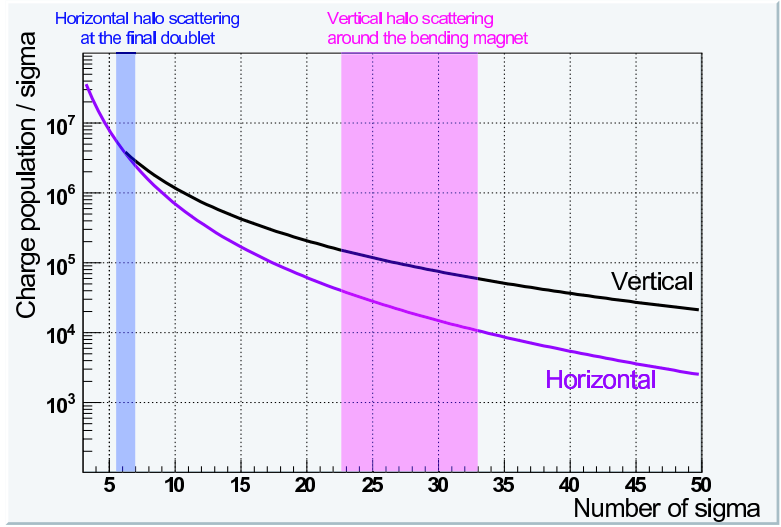}
\caption{Maximum charge density of the beam halo estimated by the halo measurement.
Blue and purple area shows the concerned region, discussed in Section \ref{sec:detector:scatpos}.}
\label{fig:detector:beamhalo-charge}
\end{center}
\end{figure}

The result of the calculation shows that 
\begin{eqnarray}
	\rho_{h1} &=& 2.2 \times 10^9 \times x^{-3.5} \qquad (\mbox{horizontal and vertical until 6 $\sigma$})
\label{eqn:detector:horzhalo}\\
	\rho_{h2} &=& 3.7 \times 10^8 \times x^{-2.5} \qquad (\mbox{vertical outside 6 $\sigma$})
\label{eqn:detector:verthalo}
\end{eqnarray}
where $x$ is the distance from the beam center as a unit of $\sigma$.
Figure \ref{fig:detector:beamhalo-charge} shows the functions.
This estimated charge population is used in the following discussions.

\subsubsection{Considerations for Beam Optics}
\label{sec:detector:scatpos}

According to the previous section, distribution of the beam halo is power functional, and
the halo expansion is proportional to the central beam size.

\begin{figure}
\begin{center}
\includegraphics[width=40em]{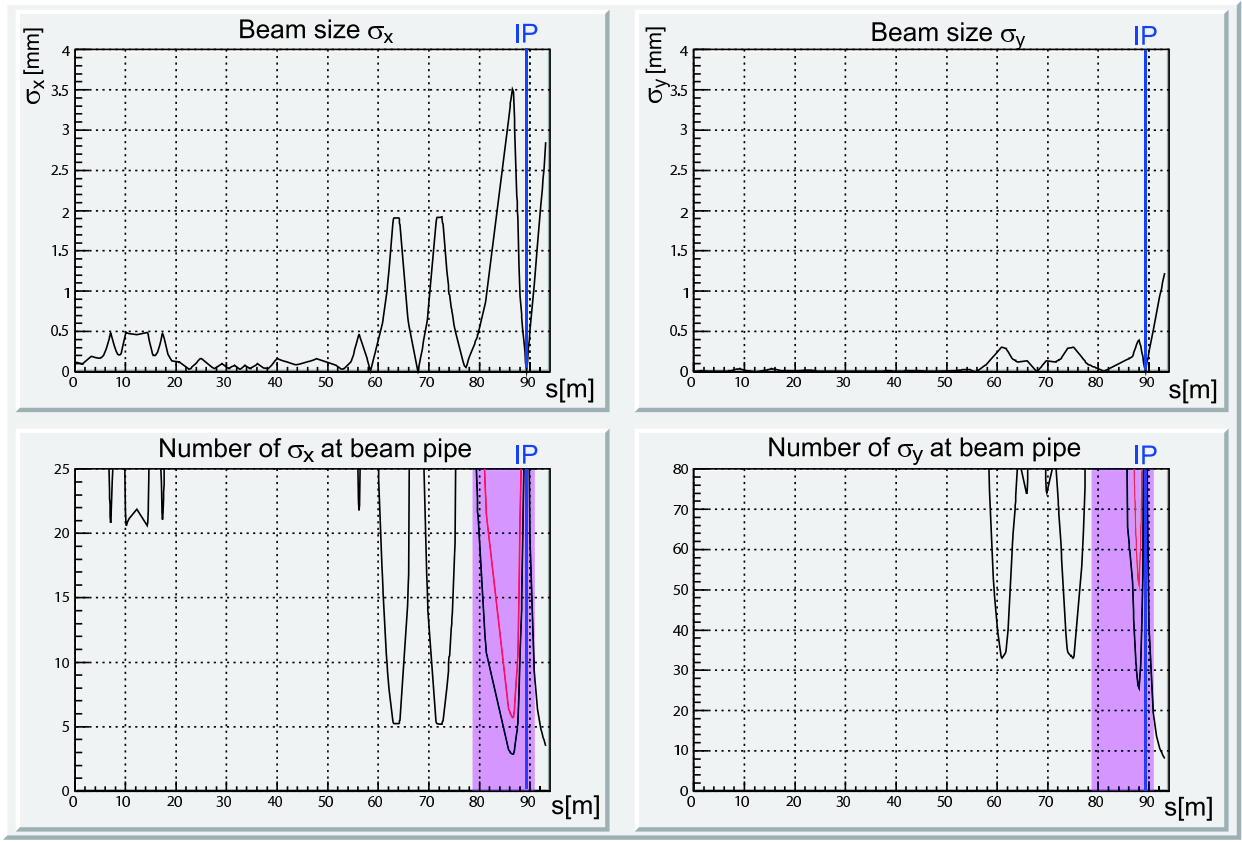}
\caption{Upper: beam size plot in the planned ATF2 line. Horizontal axis shows
$s$[m], which is a distance along the beam path from the entrance. Left plot shows horizontal beam size,
and right plot shows vertical. Lower: Number of $\sigma$ at a beam pipe of each components. Diameter of
the beam pipe is fixed to 20 mm in this plot.
The red line at the final focus section shows a case of 40 mm diameter. 
The pink area shows a straight section around the IP,
where produced background can geometrically reach the Shintake gamma detector.}
\label{fig:detector:beamoptics}
\end{center}
\end{figure}

Figure \ref{fig:detector:beamoptics} shows plots of beam parameters at each components.
Aperture diameter of the beam pipe is limited to 20 mm by
QBPM (Quadrupole Beam Position Monitor)s equipped
with every quadrupole magnet.

In the final focusing section (hatched in pink),
beam size is extremely large, especially in horizontal and
the aperture is very small with respect to that.
To avoid large amount of background produced at that section,
special QBPMs with double aperture (40 mm) will be installed
in this section.

The aperture at the final focusing magnets are still too small 
to avoid all of the beam halo even if we use 40 mm diameter QBPMs.
However, we can see sections which have smaller aperture, around $s=60-75$ m.
Most of the beam halo is cut by this section and almost no halo particles
are scattered in the final focus section.

Practically, we assume that small number of electrons may pass through the
collimators and hit the pipe at the final focus section.
The maximum number of the electrons in 5.5 - 7 $\sigma$ region
is estimated to be around $5.6 \times 10^6$ using (\ref{eqn:detector:horzhalo}), and we assume
$1.0 \times 10^6$ electrons remained and scatter with the final doublet.

Another problem exists in downstream of the IP.
The final focusing magnets strongly
focus the electron beam with large focusing angle towards the IP,
and the large focusing angle causes large beam size after the IP.
This results in the large background from around the bending magnet after the IP.

Since the aperture at the middle of the bend is around 22.5 $\sigma$,
we assume $1.0 \times 10^6$ electrons, which is calculated by
integrating (\ref{eqn:detector:verthalo}) from 22.5 to 33 $\sigma$.

The number of halo electrons assumed to produce background are much larger
than the number of signal photons.
The background must be eliminated by an appropriate photon collimator
installed in front of the gamma detector, as described in the next subsection.

\subsection{Gamma Collimator}
\label{sec:detector:collimator}

\begin{figure}
\begin{center}
\includegraphics[width=35em]{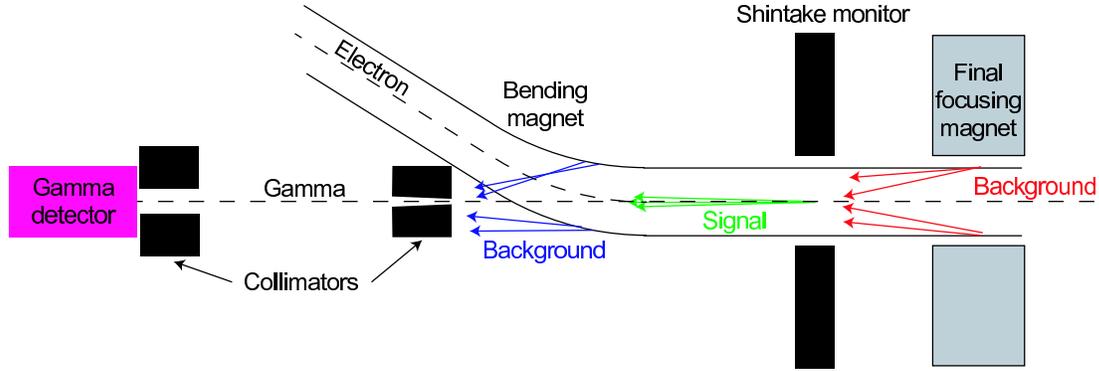}
\caption{A schematic of position of the gamma collimators.}
\label{fig:detector:gamma-collimator}
\end{center}
\end{figure}

We will install gamma collimators in front of the gamma detector to reject background.
Figure \ref{fig:detector:gamma-collimator} shows 
a schematic of the geometry. Background photons are assumed to be
produced at the final focusing magnet (shown as red arrows), and
at the bending magnet (as blue), while signal is produced
at the IP (as green).

Since distribution of the Compton signal is like a point-source and
dispersion angle is smaller than background, cone-shape collimators
can efficiently select signal photons.
Origin of the cone is set at the IP.

Opening angle of the cone is decided by comparing acceptance
for signal and background photons.
We use three opening angles in the background estimation study.
The selected opening angles are shown in Table \ref{table:detector:col-openingangle}
with calculated signal acceptance and estimated amount of background.
The collimator is located 3.0 m from the IP,
which is just behind a beam pipe of the bending magnet
crossing the gamma beam line.

Acceptance of the gamma collimator for the background photons is estimated using
a GEANT4 simulation. The acceptance for the photons from the final doublet and
that from around the bending magnet are separately investigated.

\begin{figure}
\begin{center}
\begin{minipage}[t]{.47\textwidth}
	\includegraphics[width=15em]{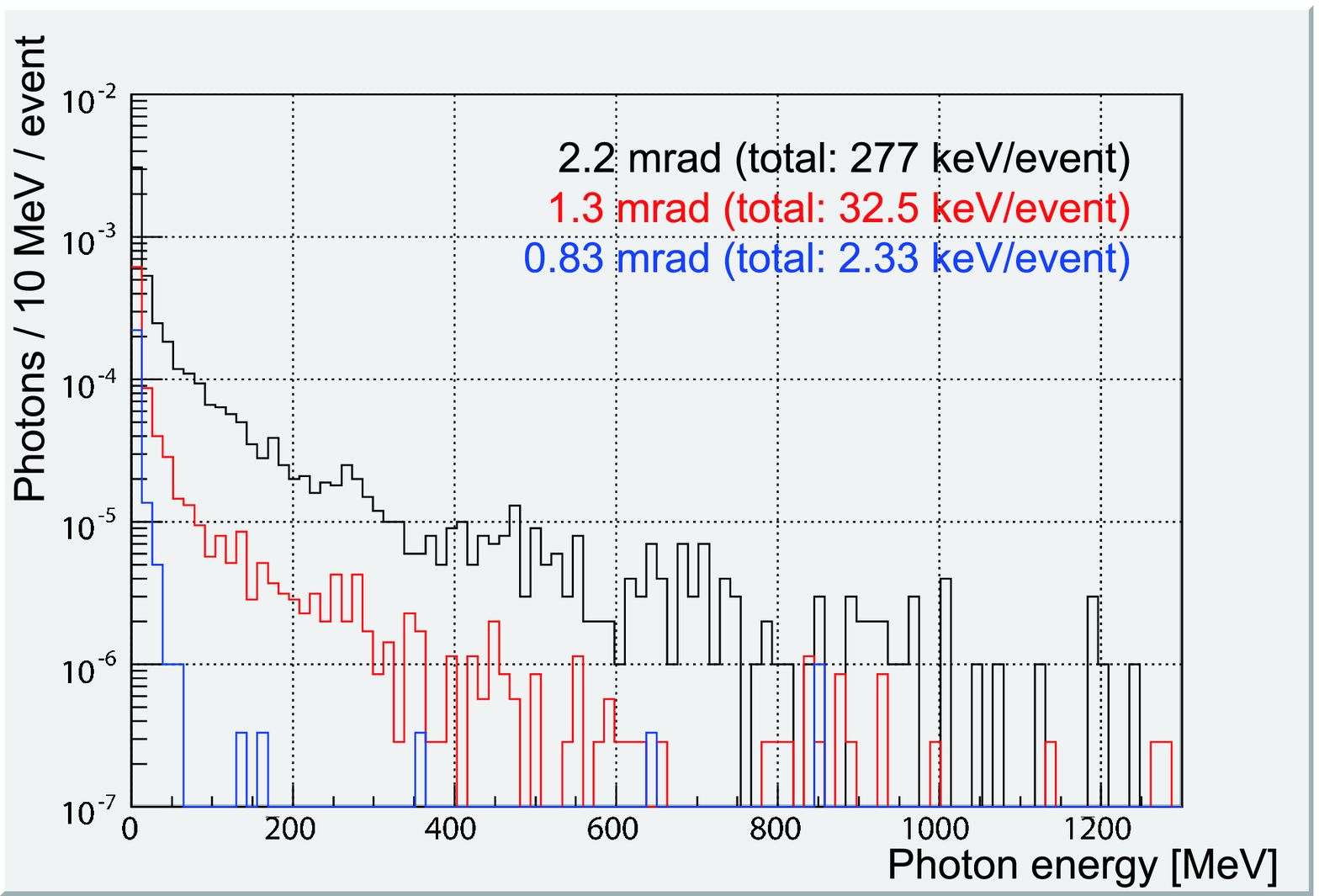}
\end{minipage}
\hfill
\begin{minipage}[t]{.47\textwidth}
	\includegraphics[width=15em]{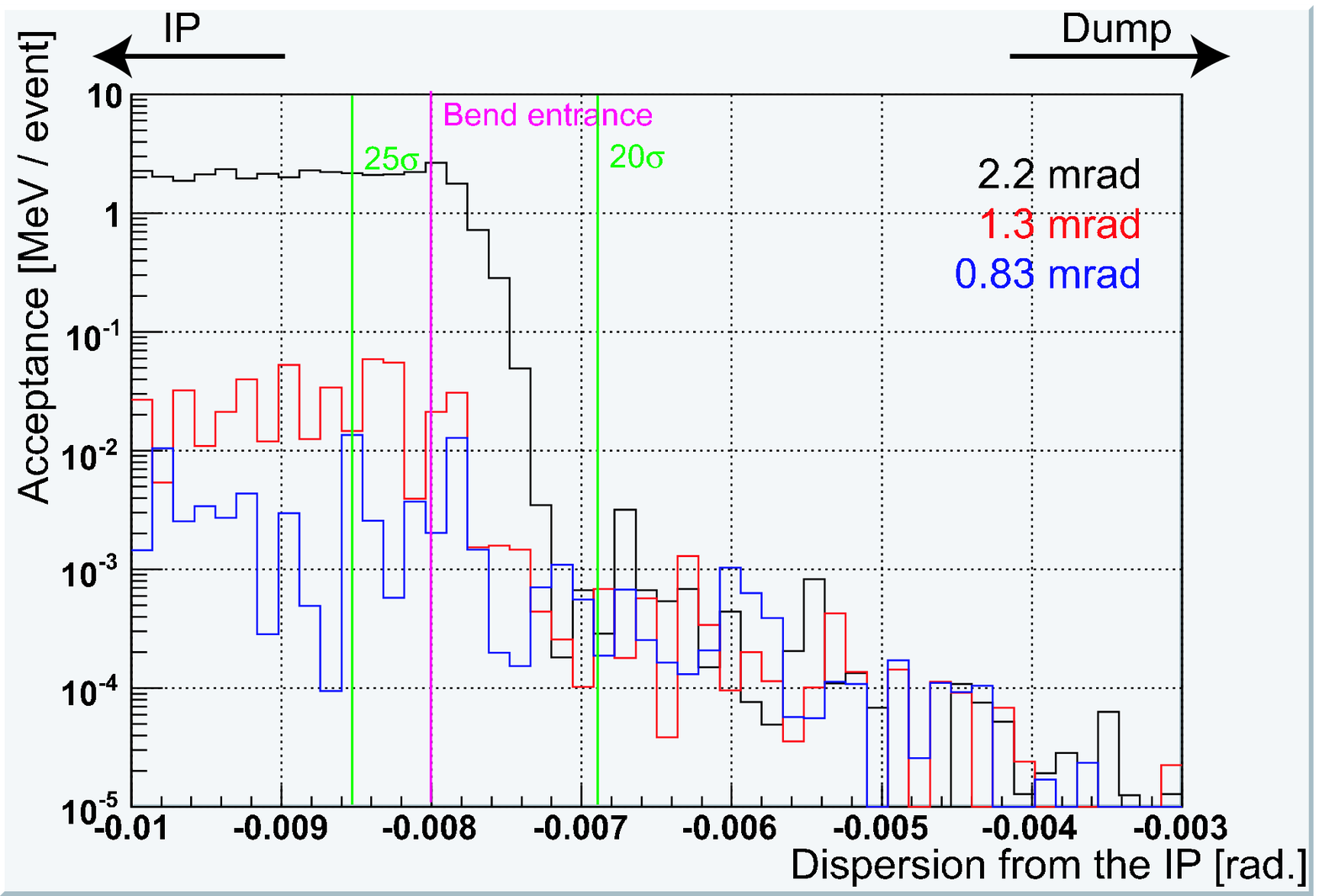}
\end{minipage}
\caption{(a)A simulation result of the acceptance of the background photons emitted
around the final doublet. Arrived energy spectra for the three collimator setups are shown.
The vertical axis stands for an arrived number of photons in a 10 MeV bin width per 1 electrons
scattering the final doublet.(b)A simulation result of the acceptance of the background photons emitted
around the bending magnet downstream the IP.
}
\label{fig:detector:scatsim}
\end{center}
\end{figure}

The result for the photons from the final doublet is shown in Fig.~\ref{fig:detector:scatsim}(a).
In this simulation, initial electrons hit the beam pipe at the final doublet
and we monitor the energy deposit of the photons at the detector, passing through
a gamma collimator. The plot shows the energy spectra of the arriving photons
passing through collimators of several apertures.

The acceptance for the background from around the bending magnet is shown in
Fig.~\ref{fig:detector:scatsim}(b). In this simulation, initial electrons come from
the IP with vertical angles from 3 to 10 mrad.~from the beam axis.
The vertical angles is projected to the horizontal axis of the plot.
The horizontal dispersion is not introduced because the horizontal beam halo is tightly
cut by the forward apertures,
while in the vertical it is cut only at 33 sigma.

The green lines show the angles corresponding to the specified number of sigma,
which is calculated from the vertical beam dispersion of the beam core (0.343 mrad.)
from the IP.
The pink line shows the entrance of the bend. Downstream the bend, the acceptance
is strongly suppressed because the electron beam is swept out from the gamma beam line.
Three histograms show the variation by the collimator aperture.

Practically the amount of the background may vary from the estimation,
and we can choose a proper aperture by monitoring the background level.

\begin{table}
\begin{center}
	\begin{tabular}{|r|r|r|r|r|r|}\hline
		Setup.	& Opening	& Signal & Final doublet	& Bending magnet & Estimated \\
		no.			& angle		&	acceptance & BG acceptance	& BG acceptance & BG photons\\ \hline\hline
		1 & 2.20 mrad & 95\% & 0.507\% & 3.64\% & 41000 \\ \hline
		2 & 1.30 mrad & 80\% & 0.060\% & 0.04\% & 1000 \\ \hline
		3 & 0.83 mrad & 60\% & 0.004\% & $<$ 0.02\% & $<$ 500 \\ \hline
	\end{tabular}
	\caption{Opening angles of assumed gamma collimators with the energy acceptance
						and the amount of background estimated by simulations.
						For estimating number of total background photons $10^6$ initial
						bremsstrahlung photons are assumed for background amount of both from final doublet
						and from bending magnet.}
	\label{table:detector:col-openingangle}
\end{center}
\end{table}

\subsection{Effects of the background}
\label{sec:bgestimation}

Since the amount of background is not modulated by verying phase of the interference fringe,
the background decreases modulation depth and can be a systematic error of the modulation measurement.
However, we can obtain pure-background signal at the detector by swiching off laser pulses.
If we set the laser pulse interleaved (which means we only fire a laser pulse for every two electron
bunches), both ``signal+background" and ``background only" data are obtained.
``Signal only" data can easily obtained by merely subtracting ``background only" data from
``signal+background" data.
By this way, systematic effect to the modulation depth is perfectly eliminated.

Statistical fluctuation to the modulation spectra is remained due to the arbitrary fluctuation
of the background amount. The rightest column of Table \ref{table:detector:col-openingangle} summarises
estimated amount of background for three collimator setups.
We should take into account two considerations:
\begin{itemize}
	\item Average background photon energy is 3.7 times larger than average signal photon energy.
		(Dependence of avarage background energy on collimator setups is small.)
	\item Energy deviation of the individual photons should be considered.
		Statistical fluctuation is 2.9 times larger than $sqrt{N}$ ($N$ is number of background photons).
\end{itemize}

Since the amount of background in Setup 1 is too large, we adopt Setup 2 (number of BG = 1000)
for resolution estimation, discussed in the next section.


\section{Performance of Shintake Monitor}

In this section resolution and accuracy of the Shintake monitor is discussed
using error estimations described in former sections.

\subsection{Sources of Statistical Fluctuations}

Several types of sources cause fluctuations
in the modulation measurements for the Shintake monitor.
In this section the amount of the fluctuations is estimated for each source.

\subsubsection{Intensity Jitters}

Intensity jitters cause random fluctuations
to each data point in the modulation spectra.
The average amount of jitters of this type
is assumed to be proportional to the signal strength.

Various error sources cause this type of error, including
laser power jitter, laser position jitter, laser timing jitter,
electron beam bunch population etc.
Some part of the intensity jitters can be canceled out using
results of individual measurements of error sources.

\begin{table}
\begin{center}
\begin{tabular}{|c|c|c|} \hline
	Error source & Jitter (uncorrected) & Jitter (corrected) \\ \hline
	Electron bunch population & 7.5\% (sample) & 1\% \\ \hline
	Laser Peak Power & 4.1\% & 3.8\% (integral correction) \\ \hline
	Laser Timing & $<$ 1\% & - \\ \hline
	Laser Spot Size & 1\% & - \\ \hline
	Laser Beam Position & 3.9\% & 1.4\% \\ \hline\hline
	Total & - & $\sim$ 4.4\%  \\ \hline
\end{tabular}
\caption{Summary of error sources of intensity jitters.}
\label{tbl:intjitter}
\end{center}
\end{table}

Table \ref{tbl:intjitter} is a list of error sources
classified to the intensity jitters. With proper correction,
the total fluctuation to each measurement point is $\lesssim$ 4.4\%.

Statistical fluctuations of measured number of photons in the detector
causes similar fluctuation to the modulation spectra, but
the amount of fluctuations of this type is not proportional to
the signal strength but proportional to $\sqrt{\mathrm{the signal strength}}$.
However, since the average number of photons is about 3000 in
expected condition, the statistical fluctuations are less than 1\%,
negligible compared to other sources.

\subsubsection{Phase Jitters}

As discussed in Section \ref{sec:fringe},
fluctuation of the laser fringe phase where the electron beam passes
is one of the critical error sources.
This type of error causes horizontal fluctuations for individual data points
in the modulation spectra.

Both fluctuations of laser fringe and electron beam position cause
this error. The amounts of errors are already estimated as 10.1 nm (fringe)
and 8.7 nm (electron) after corrections. The combined amount of error
is 13.3 nm.

\subsubsection{Background Jitters}

Fluctuation of the amount of background causes imperfect background subtraction.
By assuming that number of background photons arrived at the detector fluctuates
only by statistics, we can estimate the effects to the modulation plots.
The average number of arrival photons is estimated to be 1000 with
Setup 2 of Table \ref{table:detector:col-openingangle}.
Considering discussions in Section \ref{sec:bgestimation},
average fluctuation of background in unit of average signal energy
(53 GeV/bunch, 4400 photons/bunch $\times$ 80\% (collimator acceptance) $\times$
15.11 MeV/photon (average energy of Compton photons) ) is about 9.7\%.

Note that the background estimation is based on the worst case.
Practically the number of beam halo may be smaller and the amount of background
may also be smaller.

\subsection{Statistical Resolution of Shintake Monitor}
\label{sec:performance:statreso}

In this section, Shintake monitor is assumed to use 45 electron bunches for 
beam size measurements. This corresponds to about 1 minute (including
bunches for background subtraction), which is considered to be reasonable period.
To reduce statistical fluctuation longer measurement period is preferred,
but slow drifts (electron beam position shift, beam size shift, etc.) might
be critical for long-term measurements.

To interpret individual statistical fluctuations to resolution of the beam size measurement,
we use a toy Monte Carlo simulation. The simulation scheme is:

\begin{enumerate}
	\item Making a ``modulation plot", which is a set of measurement points.
	A measurement point simulates a signal intensity at a certain fringe phase, including
	errors (random fluctuations) discussed in the previous subsection.
	In the modulation plot, the fringe phase is changed point by point, and the scanning
	manner depends on the measurement method (discussed later).
	A simulated modulation depth is also acquired by the method using the plot.
	\item Repeating making modulation plots and obtaining a histogram of the
	simulated modulation depths. Deviation of the simulated modulation depths
	stands for the resolution for modulation measurements.
	\item Repeating those for various (true) modulation depths and measurement methods.
\end{enumerate}

As measurement methods, we use following three methods.

\begin{figure}
\begin{center}
\includegraphics[width=35em]{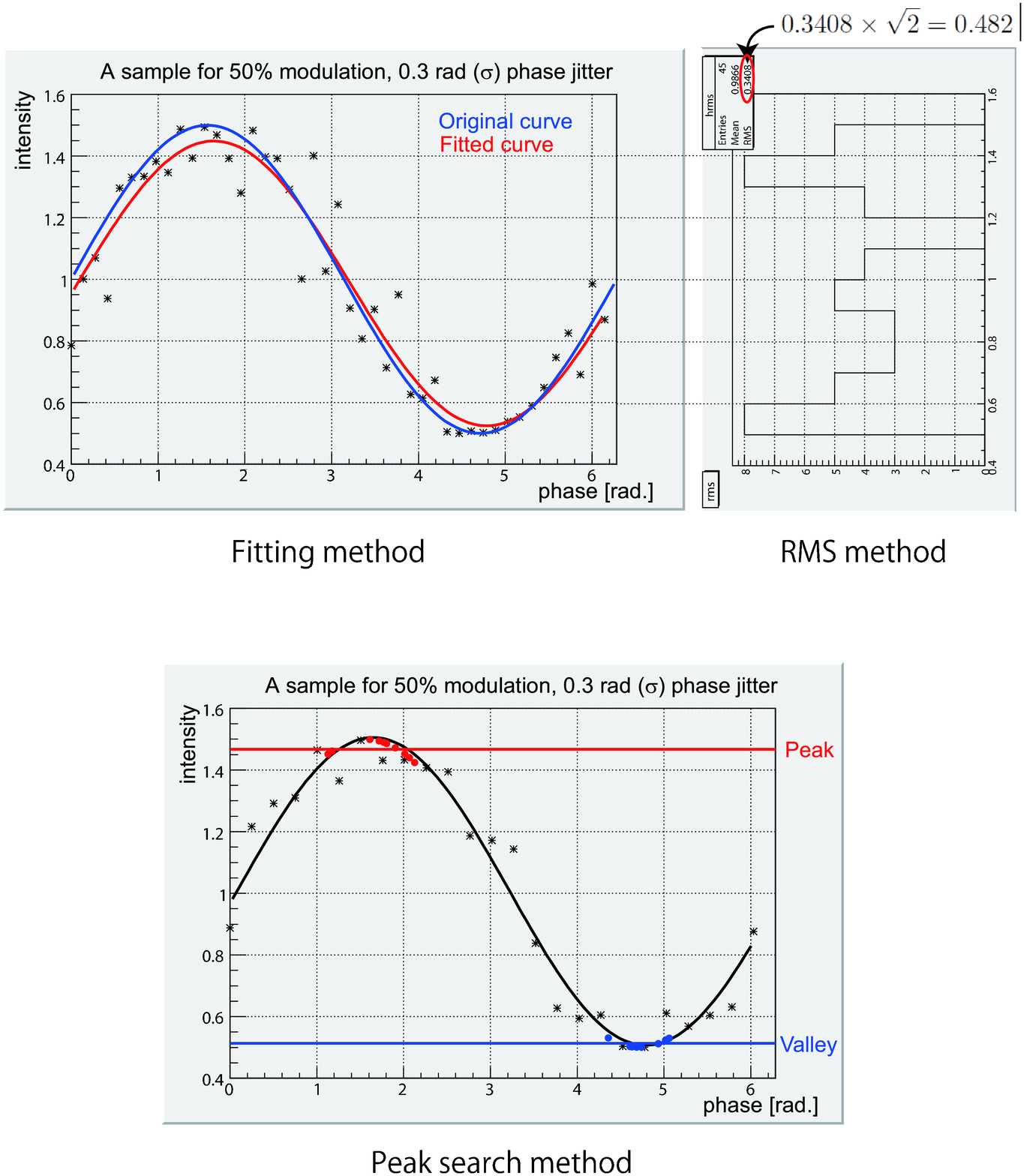}
\caption{Schematics of the measurement methods.
Top-left: the fitting method. Obtained modulation plot is fitted by a sine function
to obtain the modulation depth.
Top-right: the RMS method. Modulation points are projected, and obtain RMS value
to be interpreted to the modulation depth.
Bottom: the peak search method. First 25 points are used to obtain peak position,
and subsequent 20 pulses are used to obtain the intensity at the peak and the valley phase.
}
\label{fig:sensitivity:methods}
\end{center}
\end{figure}

\begin{itemize}
	\item Fitting Method

	The fringe phase is scanned via fixed steps over a period.
	We fit the obtained spectrum by a sine curve with free parameters of 
	amplitude, offset, and phase and the amplitude is used as a modulation depth.

	\item RMS Method

	Scanning of the fringe phase is the same as the fitting method.
	Evaluation of the modulation depth is not by fitting but merely getting RMS
	value of the scanning spectrum.

	\item Peak Search

	Peak search method consists of two stages of measurements.
	First stage consists of 25 points, obtaining the peak position of the modulation.
	In this stage the fringe phase is scanned linearly, as same as previous methods
	and fitting by a sine curve is applied immediately after scan.

	In second stage we measure the intensity at the top and the bottom of the modulation
	(each 10 points),
	using the phase information obtained in the first stage.
	Measured top and bottom data are averaged respectively, to calculate the modulation depth.
	First stage data taken at the peak search are not used in the modulation estimation.
\end{itemize}

\begin{figure}
\begin{center}
\includegraphics[width=30em]{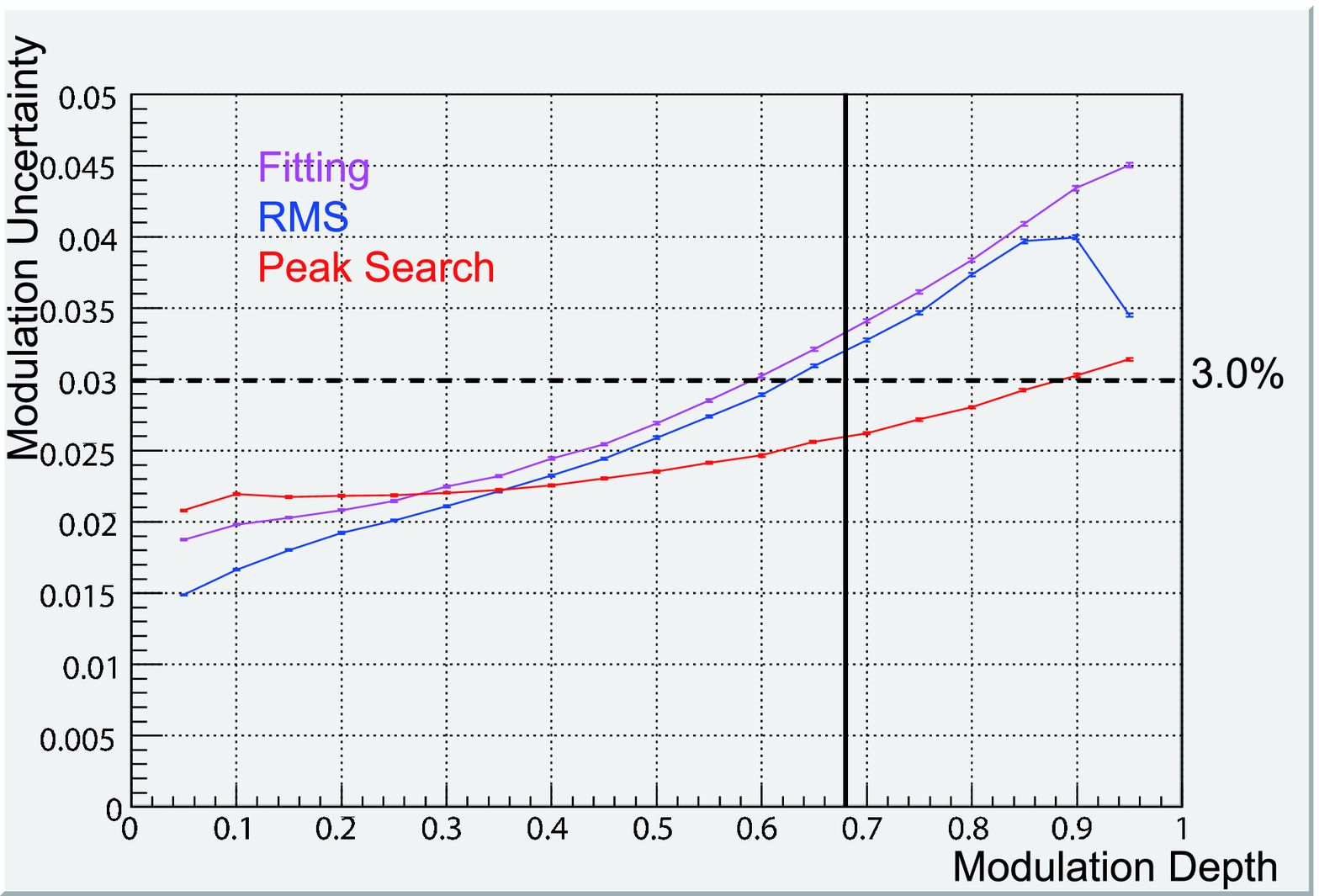}
\caption{Plot of modulation uncertainty vs.~modulation depth on each method.
Power, phase and background jitter, whose amount is described in the text, are included.
Error bars (hardly seen) stand for the error caused by limited statistics of the simulation.
The suppression of the RMS method in $M \ge 0.9$ region is due to saturations
of the measured modulation depth at $M=1$.
}
\label{fig:sensitivity:mod-reso}
\end{center}
\end{figure}

Figure \ref{fig:sensitivity:mod-reso} shows a dependence of the statistical uncertainty
of the measured modulation depth on each method.
The fitting and the RMS method give similar results,
while the RMS method gives slightly smaller uncertainty, especially for the small modulation region.
Peak search gives better result than other two methods in the range
of large modulation depth, while it is worse than the RMS method
in the modulation depth less than 35\%.

Our goal is to obtain 3\% resolution within the observable range.
With the peak search method, both of the goals can be achievable if the assumed error conditions are real.
Practically, the peak search method is used for $M>35$\% measurements
and the RMS method is used for $M \le 35$\% measurements to obtain the best resolution.

\begin{figure}
\begin{center}
\includegraphics[width=30em]{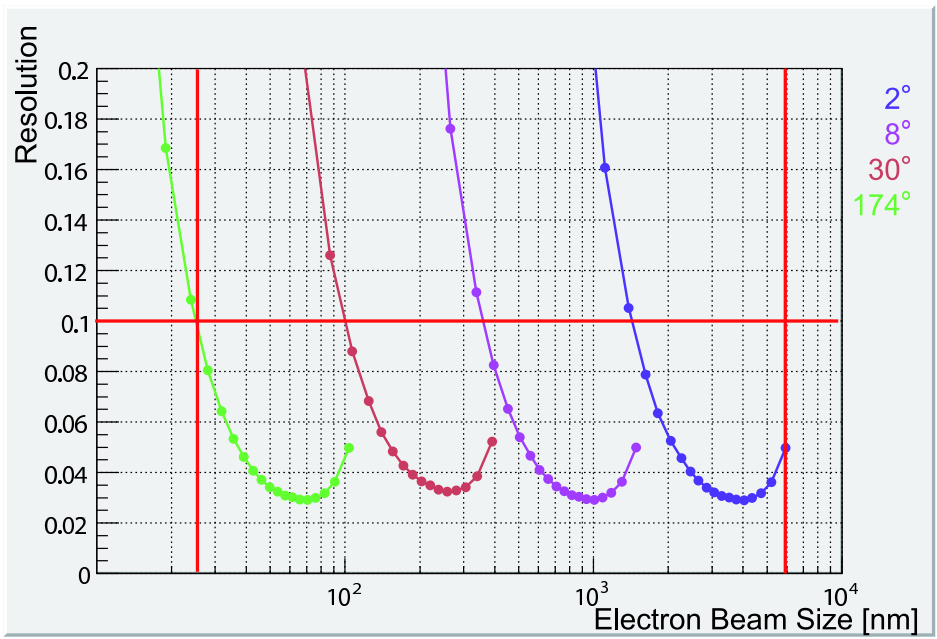}
\caption{The estimated beam size resolution.
For the 25 - 6000 nm beam size,
target resolution 10\% can be achieved using 2\degree, 8\degree, 30\degree and 174\degree
crossing angle modes.
}
\label{fig:sensitivity:all-reso}
\end{center}
\end{figure}

Figure \ref{fig:sensitivity:all-reso} shows the estimated beam size resolution using
the peak search at $M>35\%$ and the RMS method at $M\le35\%$.
From the plot, we conclude that the 10\% resolution over the observable beam size range
from 25 nm to 6 $\mu$m can be achieved.

\subsection{Systematic Errors}

For 37 nm beam size measurement, suppression of the systematic error is important.
The target accuracy of the beam size measurement is $37 \pm 2$ nm,
which corresponds to $68 \pm 2.8$\% modulation depth in the 174\degree mode.

The expected sources of the systematic effects are (1) average modulation shift
caused by the random fluctuations (discussed previously), and
(2) imperfect fringe contrast discussed in Section \ref{sec:contrast}.
We need to suppress both effects to obtain desired accuracy.

\subsubsection{Cancellation of the Average Modulation Shift}

The average modulation shift can be estimated by the same Monte Carlo simulation
as used in the previous subsection.

\begin{figure}
\begin{center}
\includegraphics[width=30em]{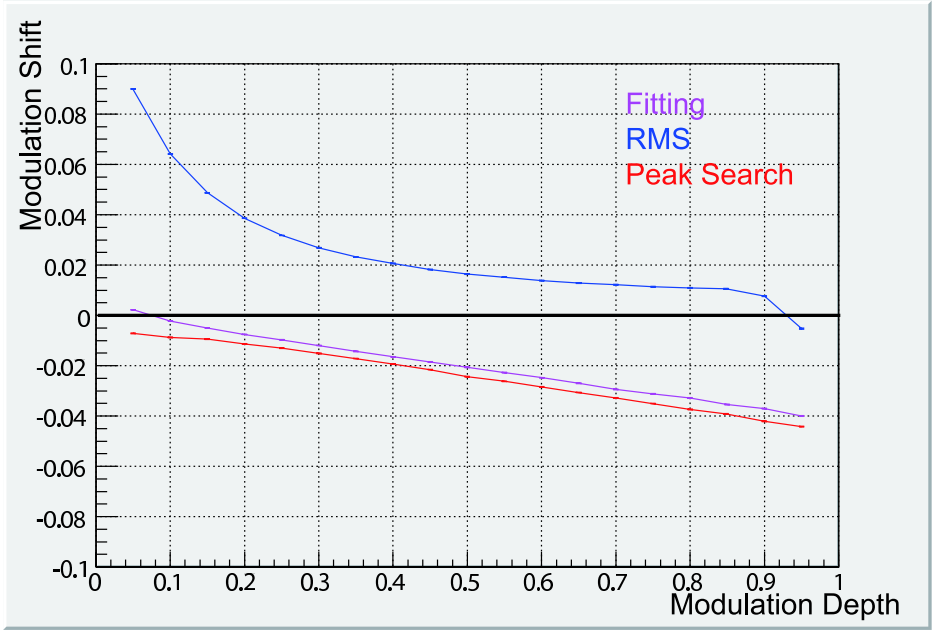}
\caption{Plot of average modulation shift on each method.
Amount of jitters is the same as Fig.~\ref{fig:sensitivity:mod-reso}.
The depression of the RMS method at $M=0.95$ is due to the saturation effect similar to
Fig.~\ref{fig:sensitivity:mod-reso}.
}
\label{fig:sensitivity:mod-ave}
\end{center}
\end{figure}

Figure \ref{fig:sensitivity:mod-ave} shows average shifts of modulation measurements
by each method.
In the fitting and peak search method,
phase jitter causes a negative shift for the obtained modulation depth,
and in the RMS method
intensity jitter causes a positive shift for the obtained modulation depth.

Since the 37 nm beam size measurement corresponds to 68\% modulation depth in the 174\degree mode,
we need to correct the shift in the peak search method.
The average modulation shift of the peak search method can be expressed as,
\begin{eqnarray}
	\delta_M &=& \frac{M}{\sqrt{2\pi}\sigma}\int^{+\infty}_{-\infty}
			\exp\left(-\frac{x^2}{2\sigma^2}\right)\cos{}xdx - M \\
					&\simeq& -\frac{\sigma^2M}{2} \quad(\sigma \ll 1)
\label{eqn:sensitivity:deltam}
\end{eqnarray}
where $\sigma$ is the average phase jitter.
If we can obtain the amount of phase jitter, the shift can be corrected.

\begin{figure}
\begin{center}
\includegraphics[width=30em]{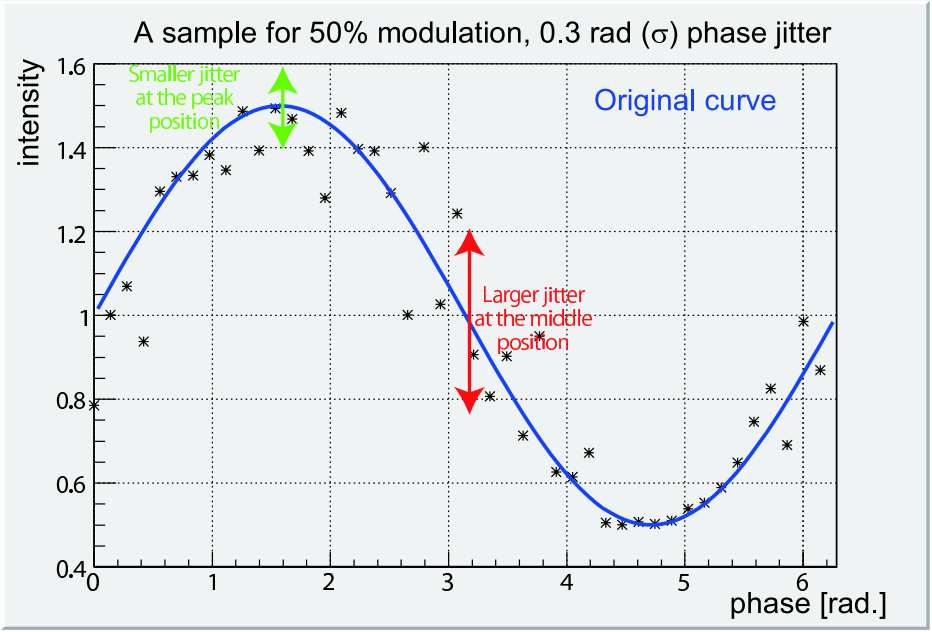}
\caption{A sample of the modulation plot with 0.3 radian phase jitter.
The power deviation with respect to the original curve is larger in the ``middle position"
than the ``peak position". Therefore, the middle position can be used to obtain the phase jitter.
}
\label{fig:sensitivity:phasejitter}
\end{center}
\end{figure}

The amount of the phase jitter can be observed by following steps.
\begin{enumerate}
	\item Acquiring background fluctuation by accumulating ``laser-off" pulses.
	\item Accumulating data with single-path light. 
					The data include ``power jitter" in addition to the background fluctuation.
	\item Measure the data with the fringe phase of the ``middle position", shown in
				Fig.~\ref{fig:sensitivity:phasejitter}.
				The power fluctuation due to the phase jitter is enhanced at the middle position.
\end{enumerate}

The required accuracy of the phase jitter for the $37 \pm 2$ nm beam size measurement
can be calculated using (\ref{eqn:sensitivity:deltam}) as,
\begin{eqnarray}
	\Delta\delta_M &\equiv& \delta_M' - \delta_M = -\frac{(\sigma'^2 - \sigma^2)M}{2} \\
	\Delta\sigma &\equiv& \sigma' - \sigma \simeq \frac{\Delta\delta_M}{M\sigma}
\label{eqn:sensitivity:deltasigma}
\end{eqnarray}
where $\Delta\delta_M$ is the accuracy of the modulation depth
and $\Delta\sigma$ is the accuracy of the phase jitter.
Assigning $M=0.68$, $\sigma=0.31$ [rad.] and $\Delta\delta_M = 0.028$, we obtain
$\Delta\sigma = 0.137$ [rad.], which is 46\% of the $\sigma$.
This resolution of the $\sigma$ can be easily obtained by using 20 bunches
for the ``middle position" measurement
(actually the resolution of the $\sigma$ is estimated to be 16\% by 20 bunch measurement).

With this correction, the systematic shift is converted to the statistical fluctuation,
caused by the resolution of the phase jitter.
The overall statistical error for the 37 nm
beam size measurement is modified from 2.6\% (see Fig.~\ref{fig:sensitivity:mod-reso})
to 2.8\%, corresponding to 2.0 nm resolution on the beam size.

\subsubsection{Treating Fringe Contrast Degradation}

As described in Chapter \ref{sec:contrast}, the fringe contrast cannot be estimated with the required accuracy.
The current estimation is $M_\mathrm{deg} < 77$\%, which has an unacceptable uncertainty, and
even the measurement result is not consistent with the estimation.

For a practical estimation of the fringe contrast, measurements with the ATF2 beam are indispensable.
As discussed in Section \ref{sec:contrast:beam}, the fringe contrast can be measured with
the accuracy of 3.54\% by the measurement of the modulation ratio.
For measurements of the 68\% modulation depth, the modulation error is $3.54\% \times 0.68 = 2.4\%$.

\subsubsection{Conclusion}

I conclude that the beam size measurement of the 37 nm ATF2 electron beam can be performed
using the Shintake monitor by,
\begin{itemize}
	\item 2.8\% modulation resolution, corresponding to 2.0 nm beam size,
	\item 2.4\% accuracy, corresponding to 1.7 nm beam size,
\end{itemize}
when the assumed conditions are achieved at the actual beam line.


\section{Summary and Prospects}


Shintake monitor is a nanometer electron beam size monitor using a laser interference fringe
formed by two split laser beams crossing at the electron beam path.
Our monitor is planned to measure 25 nm to 6 $\mu$m vertical beam size
by 2\degree, 8\degree, 30\degree and 174\degree switchable crossing angles, and 2.8 to 100 $\mu$m
horizontal beam size using laser beam scanning (laser-wire) method in the ATF2 focal point.

Concept, overall layout, critical technical issues such as phase stabilization,
beam position stabilization, fringe contrast and background of the detector are discussed in this paper.
The estimated performance is less than 10\% resolution in all measurement range 25 to 6 $\mu$m in
1 minute measurements, and less than 2.0 nm accuracy at 37 nm ATF2 design beam size.
This performance meets the requirements to achieve goals of the ATF2 experiment.
ATF2 operation will start at this October. Until summer 2009, ATF2 is planned to achieve sub-$\mu$m beam size
and the first beam size measurement by the Shintake monitor will be performed.

Shintake monitors can be useful for the real-ILC beam tuning.
For using in the ILC, several consideration should be needed.

\begin{itemize}
	\item Because the beam energy of the ILC is much larger than the ATF2, the cross section of Compton scattering
	is lower, about 1/10 of the ATF2. We need more laser beam intensity or stronger laser focusing to maintain 
	the signal strength of the gamma detector.
	\item As the peak energy of the Compton scattering photons is almost the same as beam energy, energy separation
	of signal and pipe-scattered background is not realistic. In addition, the energy of synchrotron radiation 
	photons from focusing magnets is also larger in ILC, which should be cut by some kind of shields in front
	of the gamma detector.
	\item The IP beam size of the ILC is 5.7 nm.
	For measuring 5.7 nm beam size, wavelength of the laser beam should
	be minimized. In commercially available lasers, 193 nm excimer laser may be minimum for high power pulsed beam.
	Using a 193 nm laser, 5 nm measurement is not impossible. Assuming the same resolution of the modulation depth 
	measurement as the ATF2 goal, the resolution on beam size is about $\pm$1 nm (18 \%).
	\item If the Shintake monitor is used for the IP beam size monitor in a tuning stage, it needs a space for
	the optical table at the IP. Since the ILC detector cannot be placed during the tuning with the Shintake monitor,
	a replacement structure such as a push-pull system must be needed.
\end{itemize}

\section*{Acknowledgments}

The authors thank J.~Urakawa, N.~Terunuma and ATF collaboration for useful discussions and facility support.
T.~Shintake, H.~Matsumoto and A.~Hayakawa, designers and manufacturers of the FFTB Shintake monitor
provided us useful material and comments based on their former experience.
M.~Ross, D.~McCormick, T.~Okugi and other SLAC people gave us much help for shipping FFTB Shintake monitor
components to Japan to be utilized in the ATF2 Shintake monitor.
This research was supported by Joint Japan-US Collaboration in High-Energy Phisics.




\end{document}